\documentclass[final,authoryear,3p,times,twocolumn,fleqn]{elsarticle} 

\usepackage{filecontents}

\usepackage{textcomp}

\usepackage{amssymb}
\usepackage{amsmath}
\usepackage{gensymb}
\usepackage{xcolor}
\usepackage{bibunits}

\usepackage{lineno}

\usepackage[hidelinks]{hyperref}

\journal{Earth and Planetary Science Letters}

\begin{document}

\begin{frontmatter}


\title{Scaling laws for the geometry of an impact-induced magma ocean}

\author{Miki Nakajima$^{1,2}$}
\ead{mnakajima@rochester.edu}

\author{Gregor J. Golabek$^3$, Kai W\"{u}nnemann$^4$, David C. Rubie$^3$, Christoph Burger$^5$, Henry J. Melosh$^6$, Seth A. Jacobson$^7$, Lukas Manske$^4$, and Scott D. Hull$^1$}

\address{$^1$Department of Earth and Environmental Sciences, University of Rochester, 227 Hutchison Hall, Rochester, NY 14627,  USA.\\
$^2$Department of Terrestrial Magnetism, Carnegie Institution for Science, 5241 Broad Branch Rd NW, Washington, DC 20015, USA. \\
$^3$Bayerisches Geoinstitut, University of Bayreuth, Universit\"{a}tsstrasse 30, 95440 Bayreuth, Germany.\\
$^4$Museum f\"{u}r Naturkunde, Leibniz-Institut f\"{u}r Evolutions- und Biodiversit\"{a}tsforschung, Invalidenstrasse 43, 10115 Berlin, Germany.\\
$^5$Institute of Astronomy and Astrophysics, University of T\"{u}bingen, Auf der Morgenstelle 10, 72076 T\"{u}bingen, Germany.\\
$^6$Department of Earth, Atmospheric and Planetary Sciences, Purdue University, 550 Stadium Mall Drive, West Lafayette, IN 47907, USA.\\
$^7$Department of Earth and Environmental Sciences, Michigan State University, 288 Farm Lane, East Lansing, MI 48823, USA.\\
}

\begin{abstract}
Growing protoplanets experience a number of impacts during the accretion stage. A large impactor hits the surface of a protoplanet and produces impact-induced melt, where the impactor's iron emulsifies and experiences metal-silicate equilibration with the mantle of the protoplanet while it descends towards the base of the melt. This process repeatedly occurs and determines the chemical compositions of both mantle and core. The partitioning is controlled by parameters such as the equilibration pressure and temperature, which are often assumed to be proportional to the pressure and temperature at the base of the melt. 
The pressure and temperature depend on both the depth and shape of the impact-induced melt region. A spatially confined melt region, namely a melt pool, can have a larger equilibrium pressure than a radially uniform (global) magma ocean even if their melt volumes are the same.
Here, we develop scaling laws for (1) the distribution of impact-induced heat within the mantle and (2) shape of the impact-induced melt based on more than 100 smoothed particle hydrodynamic (SPH) simulations. We use Legendre polynomials to describe these scaling laws and determine their coefficients by linear regression, minimizing the error between our model and SPH simulations. The input parameters are the impact angle $\theta$ ($0^{\circ}, 30^{\circ}, 60^{\circ}$, and $90^{\circ}$), total mass $M_T$ ($1M_{\rm Mars}-53M_{\rm Mars}$, where $M_{\rm Mars}$ is the mass of Mars), impact velocity $v_{\rm imp}$ ($v_{\rm esc} - 2v_{\rm esc}$, where $v_{\rm esc}$ is the mutual escape velocity), and impactor-to-total mass ratio $\gamma$ ($0.03-0.5$). 
We find that the equilibrium pressure at the base of a melt pool can be higher (up to $\approx 80 \%$) than those of radially-uniform global magma ocean models. This could have a significant impact on element partitioning. These melt scaling laws are publicly available on \href{https://github.com/mikinakajima/MeltScalingLaw}{GitHub} (\href{https://github.com/mikinakajima/MeltScalingLaw}{https://github.com/mikinakajima/MeltScalingLaw}).

\end{abstract}
\end{frontmatter}
%

\section{Introduction}
\label{intro}
Protoplanets experience numerous impacts as they accrete. These impacts have shaped the configuration of the solar system, given that the origins of the Earth-Moon system \citep[e.g.,][]{HartmannDavis1975, CameronWard1976}, the Pluto-Charon system \citep[e.g.,][]{McKinnon1988, McKinnon1989, Canup2005} and perhaps the Martian moons \citep[e.g.,][]{Rosenblatt2011, Craddock2011, Citronetal2015, Rosenblattetal2016, NakajimaCanup2017, CanupSalmon2018, Hyodoetal2018} can be explained by giant impacts. Additionally, the large core of Mercury \citep[e.g.,][]{Benzetal2007} and Uranus's axis tilt may have also been the result of giant impacts \citep[e.g.,][]{Safronov1966, Slatteryetal1992, Kegerreisetal2018}.

Giant impacts are not only responsible for shaping the architecture of the planetary system, but also for determining the evolving chemistry of a protoplanet. The chemical compositions of both the mantle and core of a protoplanet evolve over time as new impactor materials are added. When an impactor hits the protoplanet (target), the outer part of the mantle becomes molten and forms a magma ocean. 
Some portion of the impactor's iron core equilibrates with the ambient mantle, while the extent depends on the impactor size, velocity and impact angle \citep[e.g.,][] {DahlStevenson2010, Deguenetal2014, Landeauetal2016, Landeauetal2021, KendallMelosh2016, LhermDeguen2018}. This equilibration enriches iron melt with siderophile elements whereas lithophile elements will be preferentially partitioned into the silicate melt. The iron continues to sink to the bottom of the magma ocean and eventually merges with the target core \citep[e.g.,][]{Stevenson1990, WadeWood2005, Rubieetal2003, Rubieetal2011, Rubieetal2015}. However, if the impactor's iron core is large, it may not have time to equilibrate with the target's mantle before merging with the target's core \citep{DahlStevenson2010}. Thus, the extent of equilibration depends on the details of the impact process.

The metal-silicate partition coefficient of element $i$ is defined as, 
\begin{equation}
D^{\rm metal-sil}_{i}=C_i^{\rm metal}/C_i^{\rm sil}, 
\label{eq:D}
\end{equation}
where $C_i^{\rm metal}$ and $C_i^{\rm sil}$ are the concentrations of element $i$ in metal and silicate at equilibrium, respectively (e.g., \citealt{Rubieetal2015}). This coefficient is a function of equilibrium temperature $T_{\rm eq}$ and pressure $P_{\rm eq}$ and of other factors such as the oxygen fugacity. Conventionally, the values of $T_{\rm eq}$ and $P_{\rm eq}$ are often associated with or assumed to be proportional to the values at the bottom of a magma ocean. Based on siderophile (iron-loving) elemental abundances (e.g., Ni, Co) in the Earth mantle, early partitioning studies with piston-cylinder experiments suggest that the metal-silicate equilibration occurs at $\approx 25$ GPa, while more recent studies, including higher pressure experiments with diamond anvil cells, propose up to 55 GPa \citep[e.g.,][]{Siebertetal2013, Fischeretal2015}. 

\begin{figure*}
  \begin{center}
    \includegraphics[scale=1.3]{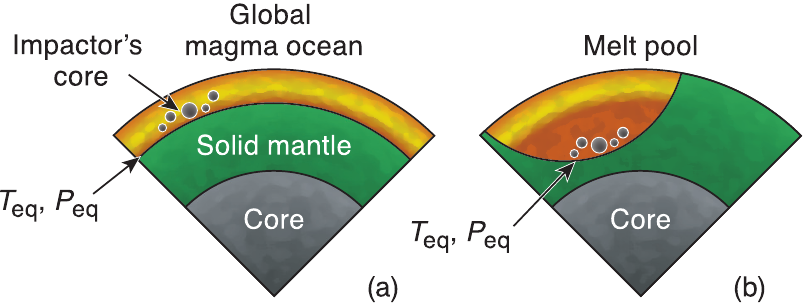}
  \end{center}
  \caption{Schematic view of (a) a global magma ocean and (b) a regionally confined melt pool. $T_{\rm eq}$ and $P_{\rm eq}$ are the equilibrium temperatures and pressures, respectively. Even if the volumes of melt are the same between the two models, a melt pool can be deeper and reach higher $T_{\rm eq}$ and $P_{\rm eq}$ than those of a global magma ocean. }
   \label{fig:meltmodels}
\end{figure*}

Conventionally, it is assumed that such an equilibration occurs in a global (radially uniform) magma ocean of equivalent volume to the melt that is generated by the impact (Figure \ref{fig:meltmodels}a). However, more realistically, an impact first produces a spatially confined melt pool (Figure \ref{fig:meltmodels}b) \citep{TonksMelosh1992, Rubieetal2015} that centers around the impact point. 
Due to isostatic adjustment this melt pool would radially spread out and become a global magma ocean over time \citep{ReeseSolomatov2006}, however, this timescale ($10^2-10^5$ years, \citealt{ReeseSolomatov2006}) is likely to be much longer than the equilibration timescale, ranging from hours (set by the turbulent mixing timescale) to months (set by the $\approx 1$ cm-sized iron droplets' sinking timescale suggested by \citealt{DahlStevenson2010}). Therefore, a melt pool is likely to be more relevant for the metal-silicate equilibration process and can provide higher pressures than a global magma ocean (Figure \ref{fig:meltmodels}). It should be noted that the geometry and pressure differences between melt pool and magma ocean diminish when the entire mantle melts. 


Insightful and extensive studies have been conducted to estimate the volume of an impact-induced magma ocean \citep[e.g.,][]{BjorkmanHolsapple1987, TonksMelosh1993, Pierazzoetal1997,PierazzoMelosh2000, ReeseSolomatov2006, BarrCitron2011, Abramovetal2012, MonteuxArkaniHamed2019}. However, some of these studies focus on head-on collisions (for which the impact angle $\theta$ is $0^\circ$; see Figure \ref{fig:angledef}a for the definition of $\theta$) because the simulations of these impacts are numerically less expensive than those of oblique impacts, which require 3D simulations, even though oblique impacts are more likely (e.g., \citealt{Shoemaker1962, Agnoretal1999}). 
Moreover, no detailed analytical models that describe how the heat is distributed within the mantle of the post-impact body for various impact angles are available. This renders challenging the prediction of the depth and geometry of an impact-induced melt pool.

Here, we have derived scaling laws for (1) the distribution of impact-induced heat within the mantle and (2) the geometry (shape) of the impact-induced melt. 
These scaling laws are expressed using Legendre polynomials and their coefficients are determined by linear regression to minimize the error between our model and impact simulations. By using these laws combined with the initial thermal profile of a planetary body, we can predict the thermal profile of the post-impact body. Moreover, once the criterion for melting is specified, the melt volume and shape of a magma ocean and melt pool can be calculated.

%
 \section{Methods}
\label{method}
\subsection{Smoothed particle hydrodynamics}
\label{sec:SPH}
We use the smoothed particle hydrodynamics (SPH) method to simulate giant impacts. SPH is a Lagrangian method and has been used for representing planetary impact phenomena (e.g.,  \citealt{Canup2004}). All the SPH particles have the same masses in a given simulation. Each SPH particle has a characteristic length scale called smoothing length. The smoothing length becomes small when an SPH particle is surrounded by neighboring particles, while the mass of the particle remains the same. This makes the density of the particle increase at a densely populated region, whereas it decreases at a sparsely populated region. The smoothing length evolves over time, but the typical scale is $\sim 200$ km. 
In SPH, the conservation equations for mass, momentum, and energy are solved simultaneously. This SPH code follows the standard implementation that uses artificial viscosity to describe the shock front (see Section 4 in \citealp{Monaghan1992}).
The SPH algorithm provides the density and internal energy at each time step, which are used to determine the pressure and sound velocity based on an equation of state (EOS). 
We use M-ANEOS as an equation of state \citep{ThompsonLauson1972, Melosh2007}, which is a semi-analytic equation of state and includes phase changes. This equation of state has been frequently used in previous impact simulations \citep[e.g.,][]{Canup2004}. The input parameters for M-ANEOS are listed in Supplementary Information (our input file is listed as ``SPH-N'' in \citealt{Stewartetal2020}). This version of M-ANEOS does not include the effect of melting, and therefore it overestimates the temperature of a material that is heated above the melting point.
The mantle and core are assumed to be dunite and iron, respectively. The initial mantle mass fractions $f_{\rm mantle}$ for both impactor and target are 0.7 (i.e. the core mass fractions are 0.3).  
Initially, the mantle and core of a body have adiabatic temperature profiles. The entropies for the mantle and core are assumed to be 3160 J/K/kg and 1500 J/K/kg, respectively, which results in approximately $\approx 2000$ K near the planetary surface and $\approx 4000$ K in the iron core at the core mantle boundary (CMB) for an Earth-sized planet. Effects of varying the initial temperatures are considered in Section \ref{sec:initT}. 
The number of SPH particles in our simulations is on the order of $10^4-10^5$, as discussed in more detail in Section \ref{sec:resolution}. The initial locations and velocities of the impactor and target are calculated as follows; (1) determining the locations and velocities of the target and impactor upon impact with the desired impact angle and velocity, and (2) calculating the trajectories of the two bodies backwards in time until the two bodies are apart by 2 Earth radii (we define this state as $t=0$ where $t$ is time). In this backward trajectory calculation, tidal deformation is ignored. For this reason, the impact angle and velocity in an SPH simulation can be slightly different from the desired values due to tidal deformation prior to the impact, but we assume that this effect is minor and it is meritorious to capture pre-impact tidal deformation because this contributes to heating.  
Our SPH code does not include material strength and the implication of this omission is discussed in Section \ref{sec:materialstrength}. 
The details of the code and settings are described in detail in our previous studies \citep[e.g.,][]{NakajimaStevenson2014, NakajimaStevenson2015}.

\subsection{Parameters for the SPH simulations}
\begin{figure*}
  \begin{center}
    \includegraphics[scale=1.3]{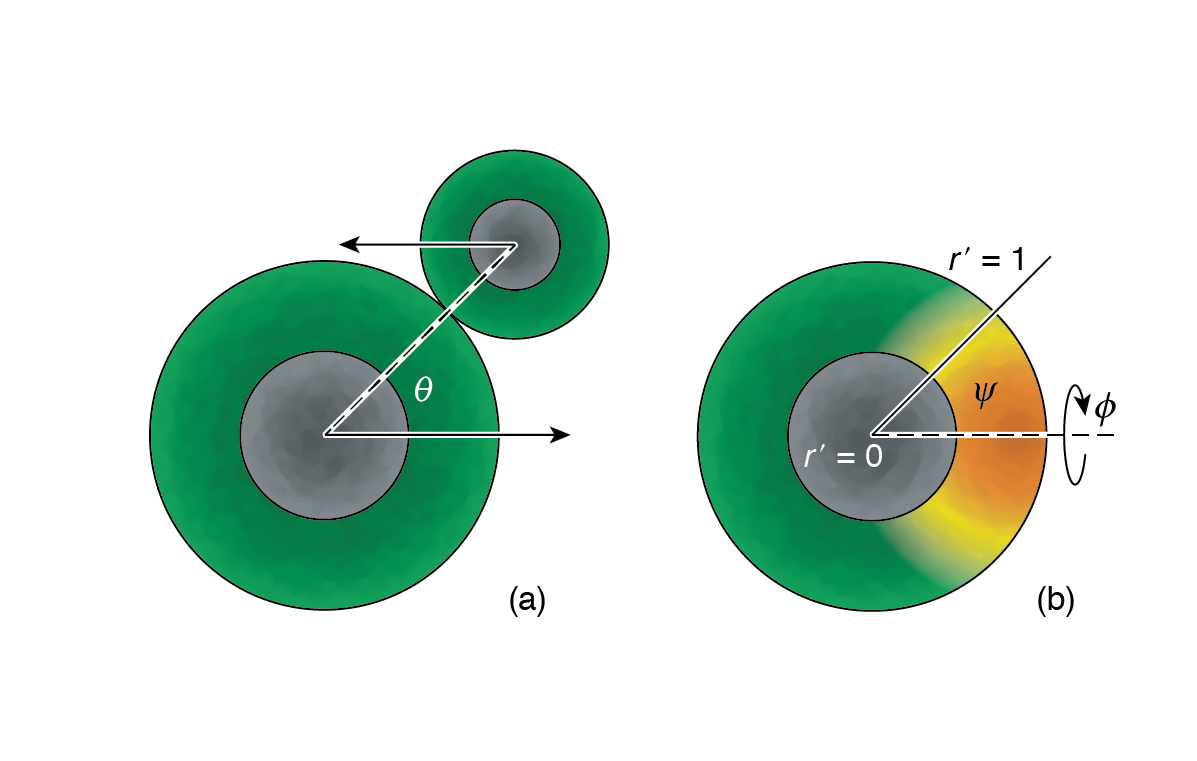}
  \end{center}
  \caption{Definition of our model parameters. (a) $\theta$ is the impact angle, where the arrows indicate the directions of motions of the impactor and the target. (b) The polar angle $\psi$ is defined to be zero where the shock-heating is most significant, which is typically close to the impact point. The heat distribution is assumed to be symmetric along the $\psi=0$ axis and therefore its dependence on the azimuth angle $\phi$ is ignored. $r'$ is the normalized radius of the post-impact body (0 is the center of the body and 1 is its surface).}
   \label{fig:angledef}
\end{figure*}

The input parameters for the SPH simulations are the impact angle $\theta$ ($0-90^{\rm o}$, Figure \ref{fig:angledef}a), total mass $M_T$ ($1M_{\rm Mars}-53M_{\rm Mars}$, where $M_{\rm Mars}$ is the mass of Mars), which is the sum of target and impactor masses, the impactor-to-total-mass ratio $\gamma$ ($0.03-0.5$), impact velocity $v_{\rm imp}$ ($v_{\rm esc} - 2v_{\rm esc}$), where $v_{\rm esc}$ is the mutual escape velocity ($v_{\rm esc}=\sqrt{2G(M_t+M_i)/(R_t+R_i)}$, where $G$, $M_t$, $M_i$, $R_t$, $R_i$ are the gravitational constant, target mass, impactor mass, target radius and impactor radius). The values of the employed parameters are listed in Table \ref{tb:parameter_definitions} and input parameters are listed in Tables \ref{tb:list1} - \ref{tb:list3}. Impacts with parameters in these ranges are expected to be common near the end of the planetary accretion stage \citep[e.g.,][]{Ward1993, Agnoretal1999, AgnorAsphaug2004}, when the impacts are largest and have the greatest influence on planetary composition. 
Figure \ref{fig:nbody} shows the distributions of these parameters in previous orbital evolution calculations \citep{Rubieetal2015}, where the typical ranges are $30^\circ \leq \theta \leq 60^\circ$, $M_T \leq 1-2 M_{\rm Mars}$, $v_{\rm imp} \leq  1.5 v_{\rm esc}$, and $\gamma \leq  0.05$.

\begin{center}
\begin{table*}[ht]
\tiny
\scalebox{1.0}{
\hfill{}
\begin{tabular}{c c c c }
\hline
Symbol & Description &  values & references \\
\hline
$\theta$ & Impact angle & 0-90$^\circ$ \\
$M_T$ & Total mass & 1 - 53 $M_{\rm Mars}$ \\
$\gamma$ & Impactor-to-total-mass ratio & $0.03 - 0.5$ \\
$M_{\rm Mars}$ & Martian mass & $6.4171 \times 10^{23}$ kg \\
$M_t$ & Target mass &  $(1-\gamma) M_T$\\
$M_i$ & Impactor mass & $\gamma M_T$ \\
$R_t$ & Target radius & $1-3 R_{\rm Mars}$ \\
$R_i$ & Impactor radius & $0.5-1.5 R_{\rm Mars}$ \\
$R^{'}$ & Radius of a planet whose mass is $M_t + M_i$ & $1-3 R_{\rm Mars}$\\
$v_{\rm imp}$ & Impact velocity  & $v_{\rm esc}$ - 2 $v_{\rm esc}$ \\
$v_{\rm esc}$ & Mutual escape velocity  & $4.2-17.5$ km/s \\
$M_{\rm mantle}$ & Post-impact mantle mass &  $0.5-1.0 M_T$  \\
$h$ & Mantle heating/total heating  & $0.7-1.0$ \\
$E_M$ & Specific energy for melting  &  $5.2 \times 10^6$ J/kg  & 1 \\
$\Delta IE$ & Internal energy gain  & $10^{29}-10^{32}$ J  & 2   \\
$KE_0$ & Kinetic energy  &  $10^{29}-10^{32}$ J &   \\
$\Delta PE$ & Change in potential energy  & $10^{29}-10^{32}$ J &   \\
$c_v$ & Specific heat &  1000 J/K/kg  & 3 \\
$f_{\rm mantle}$ & Initial mantle mass fraction &  0.7 \\
$dE_{\rm mantle}/dE$ & Fractional heating of mantle  &  0.6-1 \\
\hline
\end{tabular}}
\caption{List of key parameters used in this paper. 1: \cite{Pierazzoetal1997}, 2:  Equation \ref{eq:IE}, 3: Estimated from M-ANEOS.}
\label{tb:parameter_definitions}
\end{table*}
\end{center}


\section{Scaling law of mantle melt and heat distribution}
\label{sec:model_overview}
The SPH results are listed in Tables \ref{tb:list1}-\ref{tb:list2} for the $v_{\rm imp}=v_{\rm esc}$ cases and in Table \ref{tb:list3} for the $v_{\rm imp} \geq 1.1 v_{\rm esc}$ cases. The model name ``M'' represents the same set of initial conditions with four different impact angles (in the range of $\theta=0^{\circ} - 90^{\circ}$). ID represents a specific SPH simulation. $dE$ is the total internal energy gain of the post-impact body, $dE_{\rm mantle}/dE$ represent the fraction of the internal energy partitioned into the mantle (i.e.  $1-dE_{\rm mantle}/dE$ is the fractional energy partitioned into the core). $M_{\rm mantle}/(f_{\rm mantle} M_T)$ represents the extent of perfect or imperfect accretion of the mantle materials (if this value is close to 1, the impactor's mantle accretes into the target almost perfectly, whereas if this value is smaller than 1, some mass does not accrete into the post-impact body). The underlying assumption here is that the mantle mass fraction of the post-impact body is close to the original value, 0.7. This is an reasonable assumption in the parameter ranges we explore, but it may not be accurate if the impact velocity is much larger and the impact is catastrophic enough to change the fraction \citep[e.g.,][]{Benzetal2007}.  
$MF_A$ is the melt mass fractions of the post-impact body based on the melt criterion discussed in Section \ref{sec:heat} (Equation \ref{eq:melt}). $\sigma^\prime$ refers to the error between an SPH simulation and our model (see Section \ref{sec:heat}). $N$ is the number of SPH particles. 
Additional outputs are discussed in Section \ref{sec:additional_output_parameters}.

We describe the results of our model in terms of (1) the internal energy gain by impact in Section \ref{model1} and (2) heat distribution within the mantle in Section \ref{model2}. By combining these two sets of results, the internal energy gain and geometry of melt can be modeled as discussed in Section \ref{sec:heat}.

\subsection{Impact-induced heating}
\label{model1}
\subsubsection{SPH simulations} 
\label{sec:SPH_simulations}
\begin{figure*}
  \begin{center}
    \includegraphics[scale=0.18]{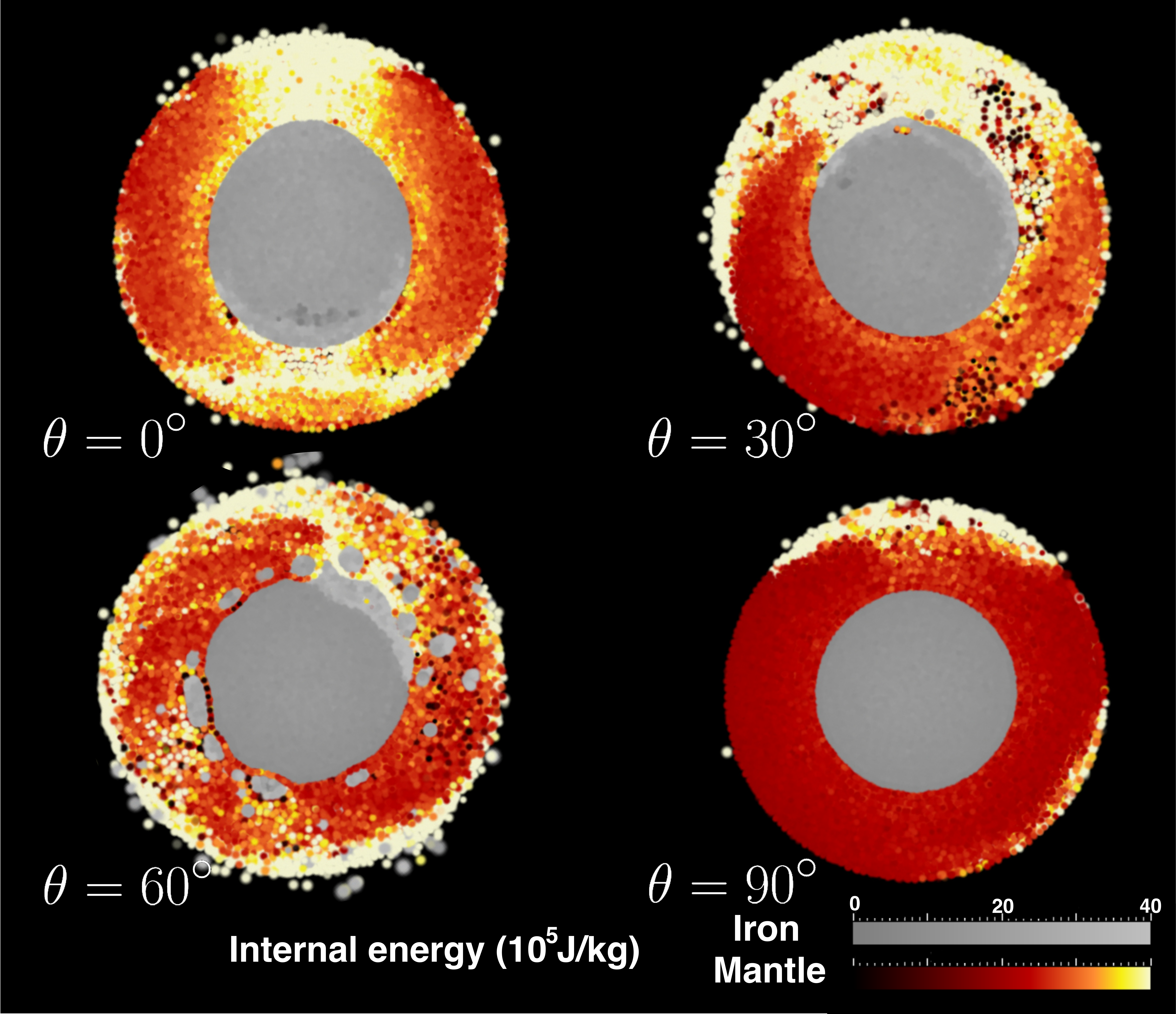}
  \end{center}
  \caption{Examples of SPH simulations (Model series M0 - see Table 2. $M_T=1 M_{\rm Mars}, \gamma=0.1, v_{\rm imp}=v_{\rm esc}$). The four panels show the results for different impact angles approximately 20 hours after the impacts. The grey and red-orange colors represent the internal energy of the iron core and silicate mantle, respectively (the values are shown in $10^5$ J/kg). $\psi=0^\circ$ is placed in the direction of 12 o'clock. The internal energy gain strongly depends on the impact angle.}
\label{fig:SPH}
\end{figure*}

Examples of our SPH simulations are presented in Figure \ref{fig:SPH} (model M0). The orange-red color map displays the gain of specific internal energy of the mantle normalized by $10^5$ J/kg and the grey color applies to iron. These snapshots clearly show that the internal energy gain depends on the impact angle $\theta$. For a head-on collision ($\theta=0^\circ$), antipodal heating is prominent due to focusing of shock waves at the opposite side of the impact point and due to deformation of the mantle. Part of the mantle at the antipode deforms significantly and expands radially upon impact. When it falls back and hits the core mantle boundary, the potential energy is converted into internal energy of the mantle. This effect is stronger at $\theta=0^\circ$ and is not clearly observed at other angles. At $\theta=30^\circ$, an impactor accretes onto the target and heats the mantle near the impact site. At $\theta=60^\circ$, the impactor hits the target twice; during the first impact, only a small portion of the impactor accretes onto the target, whereas the rest of the impactor accretes onto the target during the second impact (the so-called ``graze-and-merge collisions''). The target mantle is more uniformly heated at this impact angle (at $\theta=60^\circ$ in Figure \ref{fig:SPH}). The target's iron becomes more fragmented during the impact process. This effect is not considered in our melt model as discussed in the following sections, but it is an important effect given that this would lead to a higher extent of metal-silicate mixing compared to cases where the core is nearly intact.
At $\theta=90^\circ$, the impactor grazes the target mantle and does not accrete onto the target. Consequently, only a small portion of the target is heated.

\subsubsection{Analytical models for $KE_0$ and $\Delta PE$} 
\label{sec:dKE_dPE_analytical}
As a first step for describing the total internal energy gain $\Delta IE$, we describe the initial kinetic energy of the system $KE_0$ as
\begin{equation}
KE_0=\frac{1}{2}\frac{M_t M_i}{M_t + M_i}v_{\rm imp}^2.
\label{eq:KE0}
\end{equation} 
Assuming perfect accretion and ignoring any shape change of the post-impact body due to rotation and heating, the gain of the potential energy due to an impact, $\Delta PE$, is expressed as
\begin{equation}
\Delta PE=-\frac{3}{5}\frac{GM_t^2}{R_t}-\frac{3}{5}\frac{GM_i^2}{R_i}-\frac{GM_t M_i}{R_t+R_i} +\frac{3}{5}\frac{G(M_t+M_i)^2}{R'},
\label{eq:dPE}
\end{equation}
where $R'$ represents the radius of a body whose mass is $M_t + M_i(=M_T)$ (the mass-radius relationship between $M_T$ and $R'$ is described in Section \ref{sec:modeldetail}). The first and second terms are the gravitational binding energies of the target and impactor bodies. The third term represents the gravitational energy of the impactor body in the gravity potential of the target body, and the fourth term is the gravitational binding energy of the post-impact body under the assumption that the target and impactor perfectly merge. Equation \ref{eq:dPE} is an idealized potential energy gain assuming a perfect accretion event. The actual potential energy release can differ from this because some mass can be lost and the mass-radius relationship can change due to the temperature profile and spin of the body (see Section \ref{sec:additional_output_parameters}). Moreover, this expression does not include the effect of differentiation, which could account for up to $\sim 10 \%$ of internal energy increase (\citealt{StevensonTextbook}, D. J., online textbook). Nevertheless, $\Delta PE$ still gives a first order estimate for the potential energy change after a merging impact.

\subsubsection{Fitting models for $\Delta IE$, the mantle mass, and the fractional heating}
\label{sec:dKE_dPE_fit}
\begin{figure*}
  \begin{center}
    \includegraphics[scale=0.55]{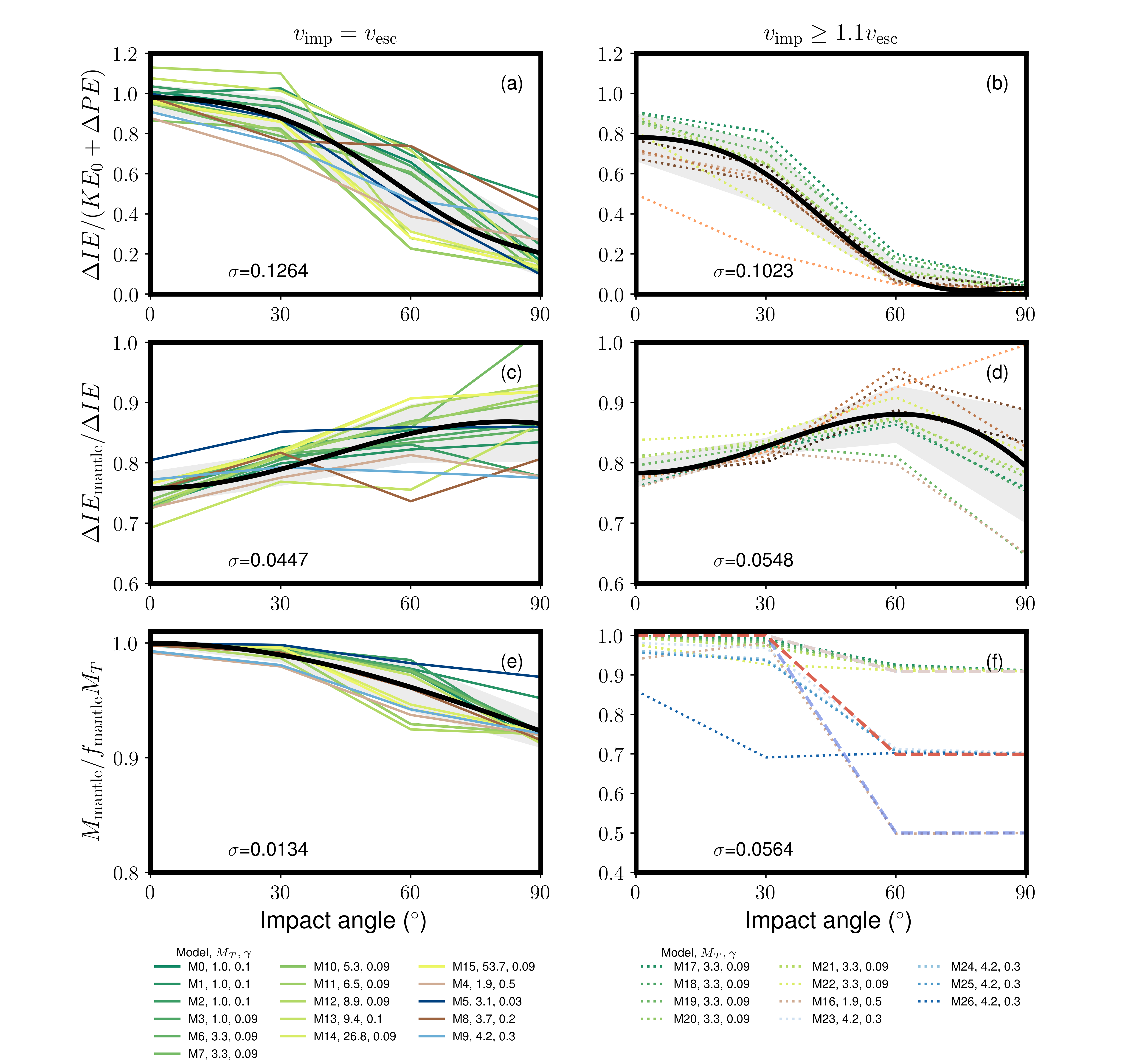}
  \end{center}
  \caption{The left panels correspond to the $v_{\rm imp}=v_{\rm esc}$ cases and the right panels correspond to the $v_{\rm imp}\geq 1.1 v_{\rm esc}$ cases.
  The empirical fits are shown with the thick black lines in panels a-e and dashed lines in panel f. SPH calculations are shown with the solid lines for the $v_{\rm imp}=v_{\rm esc}$ cases and dotted lines for the $v_{\rm imp}\geq 1.1 v_{\rm esc}$ cases. 
   (a, b) The total internal energy gain normalized by the sum of the initial kinetic energy $KE_0$ and $\Delta PE$. The colors of the solid lines represent $\gamma$ values; green-yellow lines (M0-M3, M6-M7, M10-M15, M17-M22) for $\gamma=0.09 - 0.1$, light brown (M4, M16) for $\gamma=0.5$, blue (M5) for $0.03$, brown (M8) for $0.2$, skyblue (M9, M23-M26) for $\gamma = 0.3$. (c, d) The fractional heating of the mantle with respect to the total heating. (e) $M_{\rm mantle}/f_{\rm mantle} M_T$ represent the extent of perfect or imperfect accretion at $v_{\rm imp}=v_{\rm esc}$. (f) Same as (e) but at $v_{\rm imp}  \geq 1.1 v_{\rm esc}$. The dashed lines represent Equation \ref{eq:mass_vlarge}. The legend names correspond to the models listed in Tables \ref{tb:list1}-\ref{tb:list3}. The standard deviation $\sigma$ is listed in each panel (Equation \ref{eq:error}) and the grey shadow represents the error envelope in panels (a-e).}
\label{fig:dE_fit}
\end{figure*}

We assume that the total internal energy gain $\Delta IE$ is a function of $\Delta PE$ and $KE_0$, and is expressed by the Legendre polynomials $P_l$ as
\begin{equation}
\Delta IE(\theta)=(KE_0 + \Delta PE)   \sum_{l=0}^{n_{e}} e_l P_l(\cos \theta),
\label{eq:IE}
\end{equation}
where $e_l$ are the corresponding coefficients (see Tables \ref{tb:e}-\ref{tb:e_largev}) and $n_e$ is the order of the polynomial.
The scaled internal energy gain $\Delta IE/(KE_0+ \Delta PE)$ can exceed 1 because $\Delta PE$ is an simplified energy estimate as discussed in Section \ref{sec:dKE_dPE_analytical}. 
In Figure \ref{fig:dE_fit}a, b, our best fit model is shown as a thick black line, which is modeled by sixth order Legendre polynomials ($n_{e}=6$). The lines with colors represent $\Delta IE_{\rm SPH}/(KE_0+ \Delta PE)$, where $\Delta IE_{\rm SPH}$ is the internal energy gains calculated from SPH simulations.
The left and right panels represent the $v_{\rm imp}=v_{\rm esc}$ and the $v_{\rm imp} \geq  1.1 v_{\rm esc}$ cases, respectively. The coefficients $e_l$ are determined by minimizing the error $\sigma$ (linear regression), 
\begin{equation}
\sigma = \sqrt{  \frac{1}{n} \sum_{i=1}^n \left( \frac{\Delta IE_{{\rm SPH}, i}-\Delta IE_i}{KE_0 + \Delta PE}\right)^2},
\label{eq:error}
\end{equation}
where $\Delta IE_{{\rm SPH}, i}$ is the internal energy gain from an SPH simulation whose ID is $i$ and $\Delta IE_i$ is the internal energy gain estimated with Equation (\ref{eq:IE}) for ID$=i$. $n$ is the total number of simulations we consider ($n=64$ is for $v_{\rm imp}=v_{\rm esc}$ and $n=44$ for $v_{\rm imp} \geq 1.1 v_{\rm esc}$). The colors of the lines represent different $\gamma$ values (for details, see the figure caption).
The scaled internal energy gain $\Delta IE/(KE_0+\Delta PE)$ at $v_{\rm imp}=v_{\rm esc}$ is typically larger than that at $v_{\rm imp} \geq 1.1 v_{\rm esc}$. This is because impacts at $v_{\rm imp}=v_{\rm esc}$ result in nearly perfect mergers, which can efficiently convert the impact kinetic energy and potential energy into internal energy; however, this does not always hold for cases with higher impact velocities ($v_{\rm imp} \geq 1.1 v_{\rm esc}$), which often result in hit-and-run collisions especially at large impact angles ($\theta=60^\circ, 90^\circ$) (e.g., \citealt{Asphaug2009, Gendaetal2012}). This scaled internal energy gain decreases as the impact angle increases for the same reason; at large impact angles, the kinetic and potential energies are not efficiently converted into internal energy.

In Figure \ref{fig:dE_fit}c, d, the fractional heating of the mantle with respect to the total internal energy gain, $\Delta IE_{\rm mantle}/\Delta IE$ is shown as a function of the impact angle. The fitting model $h(\theta)$ for this parameter at $v_{\rm imp}=v_{\rm esc}$ is expressed as
\begin{equation}
 h(\theta)=\sum_{l=0}^{n_g} g_l P_l(\cos \theta),
\label{eq:h}
\end{equation}
where the coefficients $g_l$ are listed in Tables \ref{tb:e} and \ref{tb:e_largev} ($n_g=2$). 
$h (\theta)$ generally increases at larger impact angles for the following reasons: an impact at a small impact angle is energetic enough to heat the core in addition to the mantle, whereas an impact at a larger impact angle tends to heat only the mantle and it is not energetic enough to heat the core. This effect can also be seen in Figure \ref{fig:SPH}, where the core is shock heated at $\theta=0^\circ$, whereas almost no strong heating occurs at $\theta=90^\circ$.

The mass of a post-impact body resulting at $v_{\rm imp}=v_{\rm esc}$ is modeled as (Figure \ref{fig:dE_fit}e)
\begin{equation}
 M_{\rm mantle}(\theta)= f_{\rm mantle} (M_t+M_i)\sum_{l=0}^{n_k} k_l P_l(\cos \theta),
\label{eq:mass_vesc}
\end{equation}
where $n_k = 1$ and the coefficients $k_l$ are listed in Table \ref{tb:e}. The best fit is shown with the thick black line in Figure \ref{fig:dE_fit}c and the corresponding coefficients $k_l$ are listed in Table \ref{tb:e} ($n_k$ = 1). 
At $\theta=0^\circ$, a target and an impactor perfectly accrete, but up to $\approx 10\%$ of the total mass, $M_T$, does not accrete at $\theta=90^\circ$ (Figure \ref{fig:dE_fit}). At $v_{\rm imp}  \geq 1.1 v_{\rm esc}$, the mantle mass of a post-impact body is not well captured by Equation \ref{eq:mass_vesc}, which assumes almost perfect accretion, because high velocity impacts tend to result in hit-and-run collisions.  At $v_{\rm imp} \geq 1.1 v_{\rm esc}$, we use the following simple imperfect accretion model (shown with the dashed lines in Figure \ref{fig:dE_fit}f), 
\begin{equation}
    M_{\rm mantle}(\theta)=
    \begin{cases}
      f_{\rm mantle}(M_t+M_i), & \text{at}\ 0^\circ  \leq \theta \leq 30^\circ  \\
      f_{\rm mantle}[M_t-M_i (\frac{\theta}{30} - 2)], & \text{at}\  30^\circ < \theta \leq 60^\circ \\
            f_{\rm mantle} M_t, & \text{at}\ 60^\circ  < \theta \leq 90^\circ. \\
    \end{cases}
    \label{eq:mass_vlarge}
\end{equation}

It should be noted that at $v_{\rm imp}=v_{\rm esc}$ and $\theta=90^\circ$, our scaled internal energy model (Equation \ref{eq:IE}) is underestimated in some cases. An impactor hits the surface of the target and continues to orbit around the target and eventually hits the target again. However, some of the SPH simulations are stopped before an impactor comes back because we only run simulations up until $\approx 20-25$ hours, when the effect of numerical viscosity becomes non-negligible \citep{Canup2004}. This can be seen in Figure \ref{fig:dE_fit}e, where some impactors accrete into targets while others do not (e.g., Models M10 and M11). We expect that this could lead to up to $\approx 10\%$ error in the internal energy given that the standard deviation at $\theta=90^\circ$ is 0.115 (see the grey error envelope in Figure \ref{fig:dE_fit}a).

\subsection{Distribution of heat}
\label{model2}
The heat distribution within a planetary mantle is also modeled with Legendre polynomials.  
We define the spatial heat distribution function $F (r', \psi, \phi)$ as
\begin{equation}
F (r', \psi, \phi) = F(r', \psi) \equiv\frac{\Delta U(r', \psi)}{\Delta \bar{U}},
\label{eq:P0}
\end{equation}
where $r'$ is the normalized radius (0 at the center and 1 at the planetary surface), $\psi$ is the colatitude, and $\phi$ is the azimuth (see Figure \ref{fig:angledef}b). We assume that the heat distribution is symmetric along the pole and therefore the model does not depend on $\phi$. This assumption is not accurate when $\theta>0^\circ$. Nevertheless, this angle dependence is relatively weak, so we ignore it in this model (see Section \ref{sec:modeldetail}  and Figure \ref{fig:xy_xz} for further discussion). 
$\psi$ is defined as zero where the impact-induced heating is maximum, which often coincides with the impact point.
$\Delta U(r', \psi)$ is the specific internal energy gain at $r'$ and $\psi$. $\Delta \bar{U}$ is the average specific internal energy gain of the mantle. 
Here, $F(r', \psi)$ is assumed to be 
\begin{equation}
F(r', \psi) = \sum_{m=-2}^{2} \sum_{l=0}^{2} c_{l+3(m+2)} r'^{m} P_l (\cos \psi).
\label{eq:F}
\end{equation}
This expression requires $5 \times 3 = 15$ coefficients $c$, which are determined as follows.
We divide an SPH simulation output into $8 \times 12$ segments as a function of radius and angle, $r'_i- \frac{1}{2}\Delta r \leq r'< r_i'+ \frac{1}{2}\Delta r'$ and $ \psi_j - \frac{1}{2} \Delta \psi \leq  \psi < \psi_j + \frac{1}{2} \Delta \psi $ ($\Delta r' \approx 0.5/8 = 0.0625$ and $\Delta \psi = 180^\circ/6 =30^\circ$). For this, we calculate the average of $\Delta U(r', \psi)/\Delta \bar{U}$ of the hemisphere at each segment (one hemisphere covers $-90^\circ \leq \phi <90^\circ$ and the other covers $90^\circ \leq \phi <270^\circ$, where $\phi=0^\circ$ and $\phi=180^\circ$ are parallel to the impact velocity vector).
To define the location of $\psi=0$, we calculate the averaged internal energy gain at each $\psi$ segment ($ \psi_j - \frac{1}{2} \Delta \psi \leq  \psi < \psi_j + \frac{1}{2} \Delta \psi $ and $0.55 \leq r' < 1$), and we identify $\psi$ at which the internal energy gain is maximum and set it as $\psi=0$.
The 15 coefficients in Equation \ref{eq:F} are determined by minimizing the error between the model and the averaged internal energies in all the segments (the 15 coefficients are listed in Table \ref{tb:coefficients0}).
It should be noted that the coefficients are determined using all the SPH simulations, including both the $v_{\rm imp}=v_{\rm esc}$ and  $v_{\rm imp}\geq 1.1v_{\rm esc}$  cases. 
We also explored different orders ($r'^{-3}, r'^{3}, P_3(\cos \psi)$), but their effects were limited and therefore we did not include these terms in the model. 

\begin{figure*}
  \begin{center}
    \includegraphics[scale=0.45]{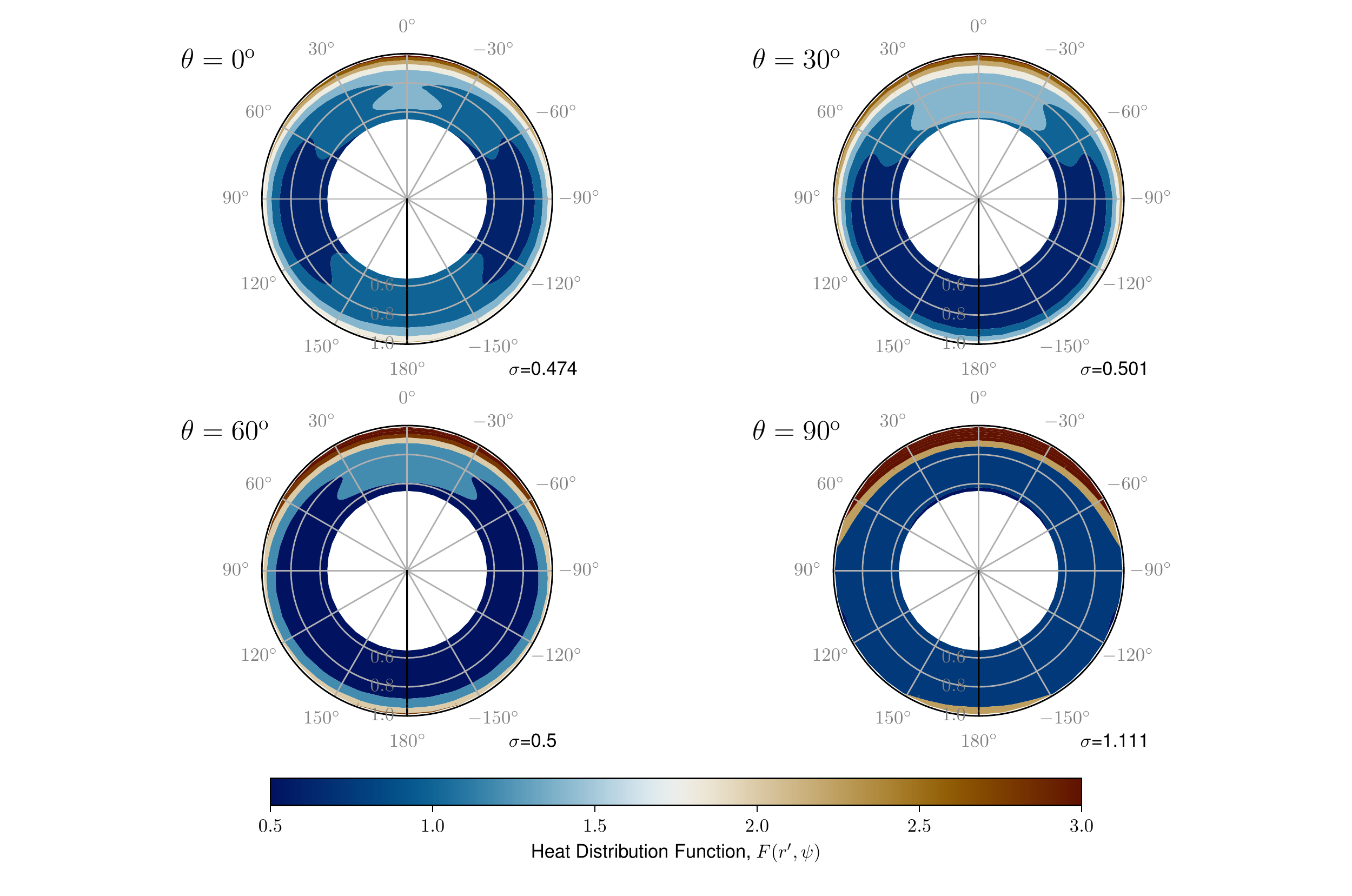}
  \end{center}
  \caption{Normalized heat distribution model $F(r', \psi)$ (Equation \ref{eq:F}) for four different impact angles. The internal energy is normalized by the averaged energy of the system. The error is calculated at each impact angle.}
  \label{fig:Heatmodel}
\end{figure*}

$F(r', \psi)$ is shown in Figure \ref{fig:Heatmodel}. The antipodal heating is well captured at $\theta=0^\circ$, which is not clearly seen in the case of the other impact angles. Interestingly, the mantle is heated more uniformly at small impact angles ($\theta=0^\circ, 30^\circ$) than at larger impact angles ($\theta=60^\circ, 90^\circ$). This finding may seem counter-intuitive, but this can be explained given that an impact with a small impact angle often results in accretion, which is an efficient way to heat the whole mantle, whereas an impact with a large angle heats only the near surface regions of the target body.

\subsection{Comparison between our heat distribution model and SPH}
\label{sec:heat}
\begin{figure*}
  \begin{center}
    \includegraphics[scale=0.5]{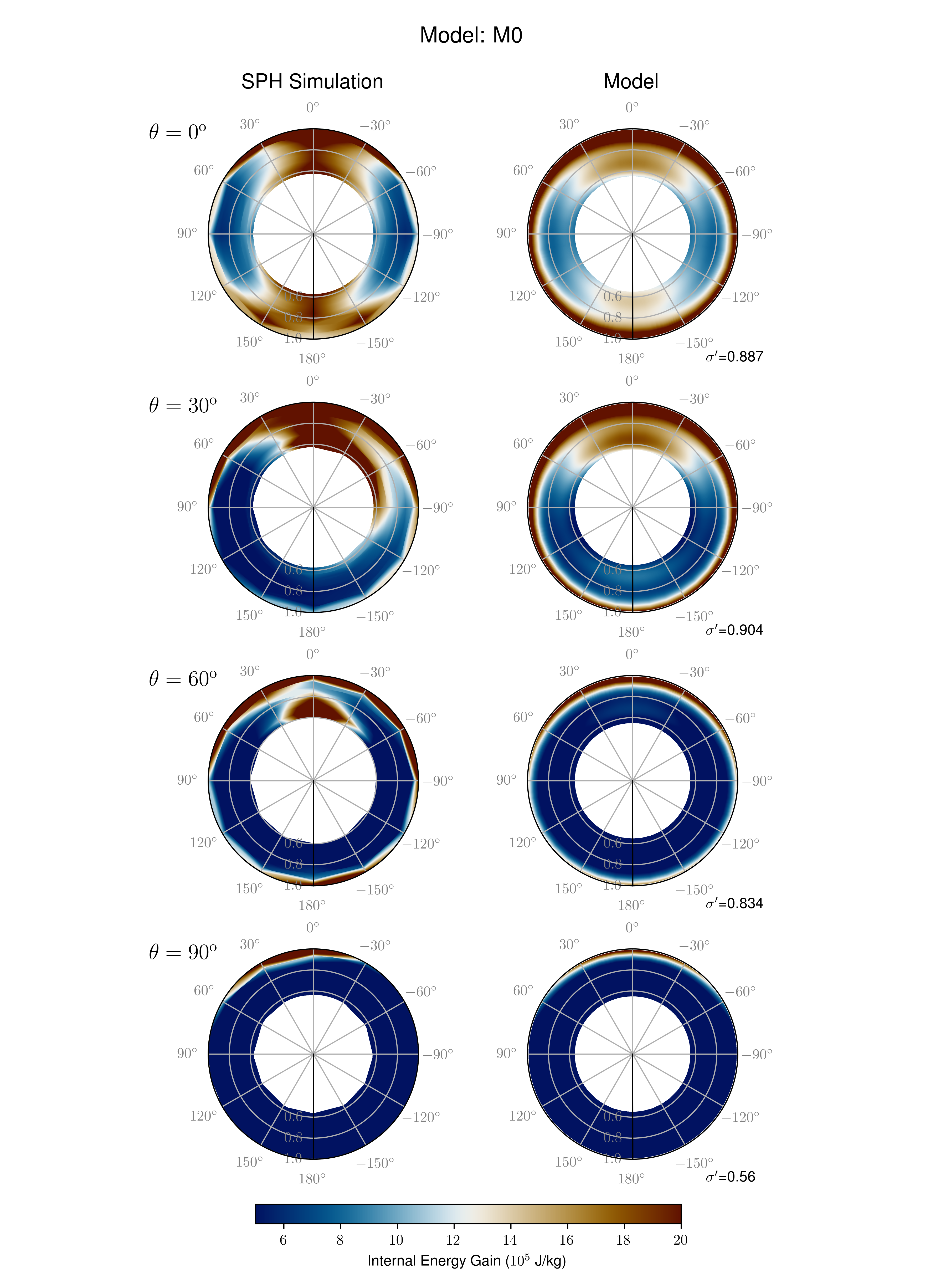}
  \end{center}
  \caption{Comparison between SPH results (left) for M0 model series (Table \ref{tb:list1}) and the scaling law (right) for (top row) $\theta=0^{\circ}$, (second row) $30^{\circ}$, (third row) $60^{\circ}$, (bottom row) $90^{\circ}$, respectively. The input parameters are $M_T=1 M_{\rm Mars}, \gamma=0.1, v_{\rm imp}=v_{\rm esc}$.
  The color contour represents the internal energy normalized by $10^5$ J/kg after the system reaches its equilibrium (typically within 10 hours). The heat distribution of the mantle based on the SPH simulation is not smooth at several places partly because SPH data are segmented and partly because the internal energy, especially near the core-mantle boundary, is smaller than the contour intervals.  The standard deviation $\sigma^\prime$ is also normalized by $10^5$ J/kg.}
  \label{fig:M0d}
\end{figure*}

\begin{figure*}
  \begin{center}
    \includegraphics[scale=0.6]{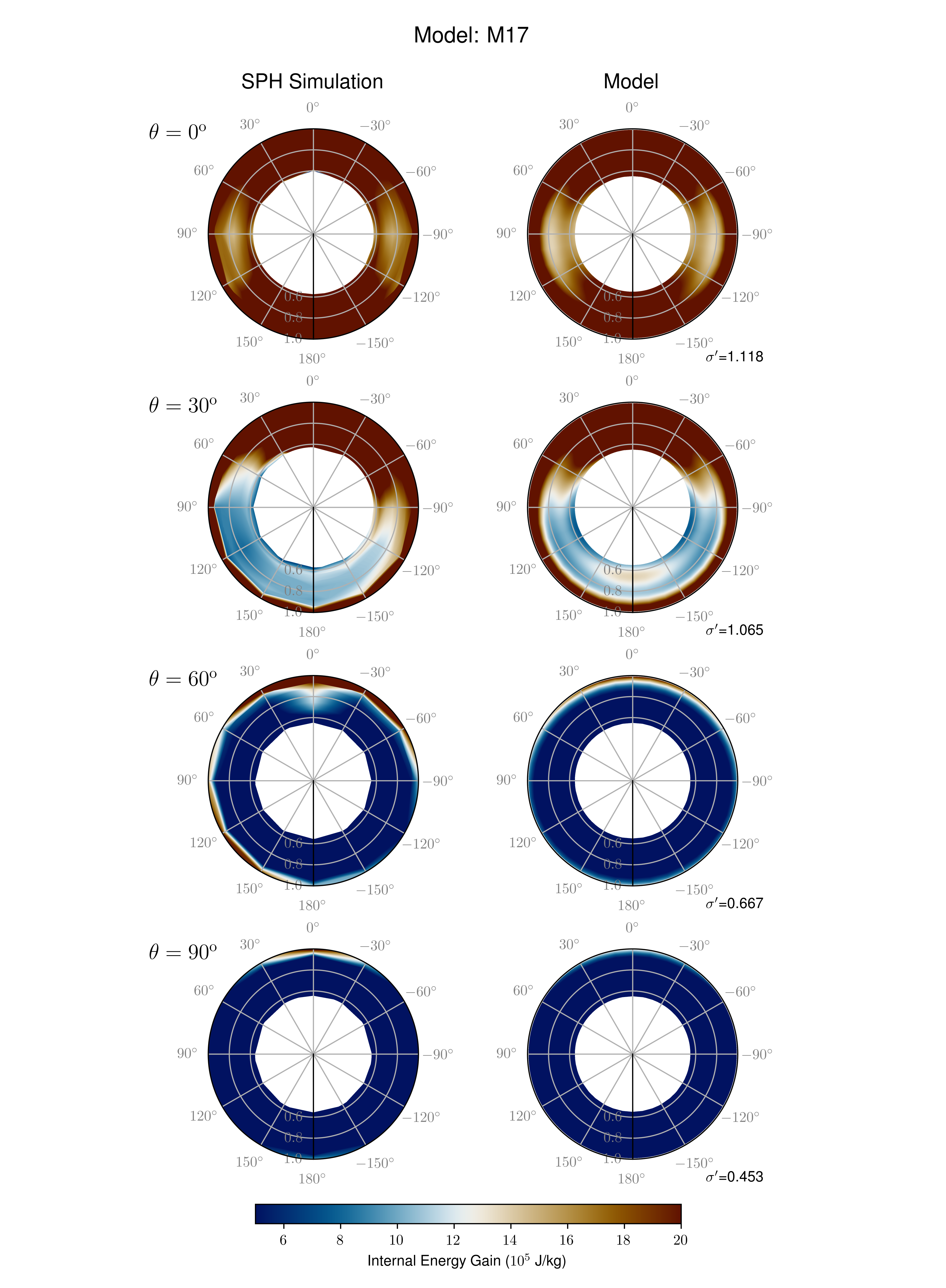}
  \end{center}
  \caption{SPH results for model M17 (Table \ref{tb:list3}). The color scheme is the same as the one in Figure \ref{fig:M0d}.  The input parameters are $M_T=3.25 M_{\rm Mars}, \gamma=0.091, v_{\rm imp}=1.2 v_{\rm esc}$. The standard deviation $\sigma^\prime$ is normalized by $10^5$ J/kg.}
  \label{fig:M18d}
\end{figure*}

$\Delta \bar{U}$ (Equation \ref{eq:P0}) is calculated by averaging the internal energy gains of SPH particles. We assume that $\Delta \bar{U}\approx \Delta IE(\theta)$ where $\Delta IE(\theta)$ is a model fit (Equation \ref{eq:IE}). This leads to 
\begin{equation}
\Delta U(r',\psi, \theta)=F(r', \psi) \Delta IE(\theta).
\label{eq:dU}
\end{equation}
Figures \ref{fig:M0d} and \ref{fig:M18d} show the internal energy gains calculated from the SPH simulations (left) and model $\Delta U$ (right) for the models $M0$ and $M17$, respectively. Here, the internal energy gain is normalized by $10^5$ J/kg. The standard deviation $\sigma^\prime$ is described as
\begin{equation}
\sigma' = \sqrt{\sum_{i=1}^{n_{r'}} \sum_{j=1}^{n_\psi} \frac{1}{n_{r'} n_\psi } \left[\Delta IE_{\rm SPH}(r'_i,\psi_j) - \Delta IE(r'_i,\psi_j)\right]^2},
\label{eq:L2}
\end{equation}
where $n_{r'}=8$ and $n_{\psi}=12$. 
In general, the overall trend is captured in our model; at $\theta=0^\circ$, the mantle is extensively heated near the impact and antipodal site, whereas the mantle remains colder at $\psi=90^\circ$ and $-90^\circ$. At $\theta=30^\circ$, the mantle on the hemisphere close to the impact ($|\psi|<90^\circ$) is significantly heated, whereas the other side of the hemisphere ($90^\circ<|\psi|<180^\circ$) is much less shock-heated. At $\theta=60^\circ$ for M0, a portion of the mantle near the core mantle boundary is highly shock heated in the SPH simulation. This is because part of the mantle is locally heated while the impactor's core sinks to the bottom. This is not captured in the model. At $\theta=90^\circ$, our model underestimates impact-induced heating in models M0 and M17 primarily because our $\Delta IE$ model also underestimates heating (Figure \ref{fig:dE_fit} a,b). Our model works well at small $\gamma$, but it does not work as well at $\gamma=0.5$ where an impactor is as large as the target (Figure \ref{fig:M4d}). This is because this impact is more energetic than cases with smaller $\gamma$ and our model underestimates the extent of heating.

\subsection{Effect of the initial temperature}
\label{sec:initT}
\begin{figure*}
  \begin{center}
    \includegraphics[scale=0.4]{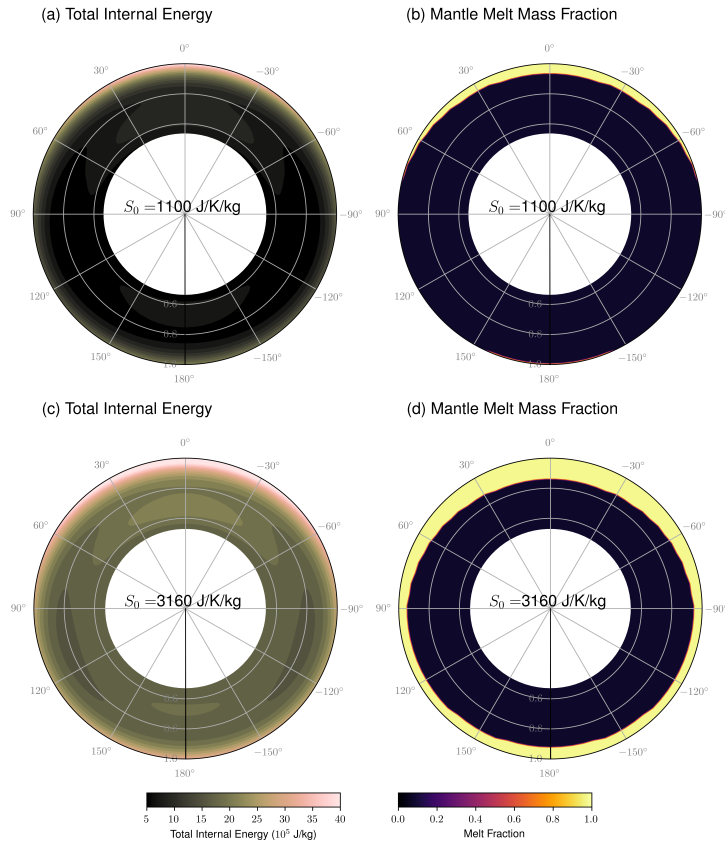}
  \end{center}
  \caption{Effect of initial temperature is considered for model M0 at $\theta=60^\circ$. (a) Total internal energy assuming  initial entropy $S_0=1100$ J/K/kg (this corresponds to a surface temperature of $\approx 300$ K). The total internal energy is calculated as a summation of the pre-impact internal energy and the internal energy gain calculated in Equation \ref{eq:dU}. (b) Melt mass fraction. When the mantle temperature is above the melt temperature, this value is 1, otherwise the value is 0 (no partial melting is considered). (c) is the same as (a) except $S_0=3160$ J/K/kg (this corresponds to a surface temperature of  $\approx 2000$ K). (d) is the same as (b) except $S_0=3160$ J/K/kg.}
  \label{fig:initialT}
\end{figure*}

In this section, we explore the geometry of a magma ocean by considering two different initial temperature profiles of the mantle.
In Figure \ref{fig:initialT}, the panels (a) and (c) represent the total internal energy and the panels (b) and (d) represent the melt fraction for M0 at $\theta=0^\circ$. The pre-impact mantle entropy $S_0$ is 1100 J/K/kg for (a) and (b) (the surface temperature of $\approx 300$ K) and 3160 J/K/kg for (c) and (d) (the surface temperature of $\approx 2000$ K). These isentropic temperature profiles are calculated using M-ANEOS (see Section \ref{sec:modeldetail}).
The total internal energy is calculated as the sum of the initial internal energy and internal energy gain using Equation \ref{eq:dU}.
We define a portion of the mantle is molten if the local temperature exceeds the melting temperature, $T_{\rm melt}$  \citep{Rubieetal2015}, 
\begin{gather} 
P \leq  24\ {\rm GPa}: T_{\rm melt} { \rm [K]} = 1874 + 55.43 P - 1.74 P^2 \nonumber \\  
+ 0.0193 P^3 \nonumber \\ 
P > 24\ {\rm GPa}: T_{\rm melt} { \rm [K]} = 1249 + 58.28 P - 0.395 P^2 \nonumber \\  
+ 0.0011 P^3 ,
\label{eq:melt}
\end{gather}
where $P$ is the pressure in GPa. The local temperature is calculated assuming $u=c_v T$, where $c_v$ is the specific heat and $T$ is the temperature. In panels (b) and (d), the yellow regions represent melt and black regions represent solid mantle. Pressure is calculated with M-ANEOS assuming that the planet is in a hydrostatic equilibrium and isentropic. 
No partial melting is considered. As shown in Figure \ref{fig:initialT}, different initial temperatures affect not only the melt volume, but also the shape of the melt. In panel (b), the melt pool is confined to one hemisphere ($|\psi| <90^\circ$), whereas the melt volume is larger in panel (d) and the melt is globally distributed. It should be noted that the M-ANEOS used here does not consider solid-liquid transition and therefore this criterion overestimates the melt volume.

\subsection{Magma ocean depth and the corresponding pressure}
\label{modepth}
\begin{figure*}
  \begin{center}
    \includegraphics[scale=0.5]{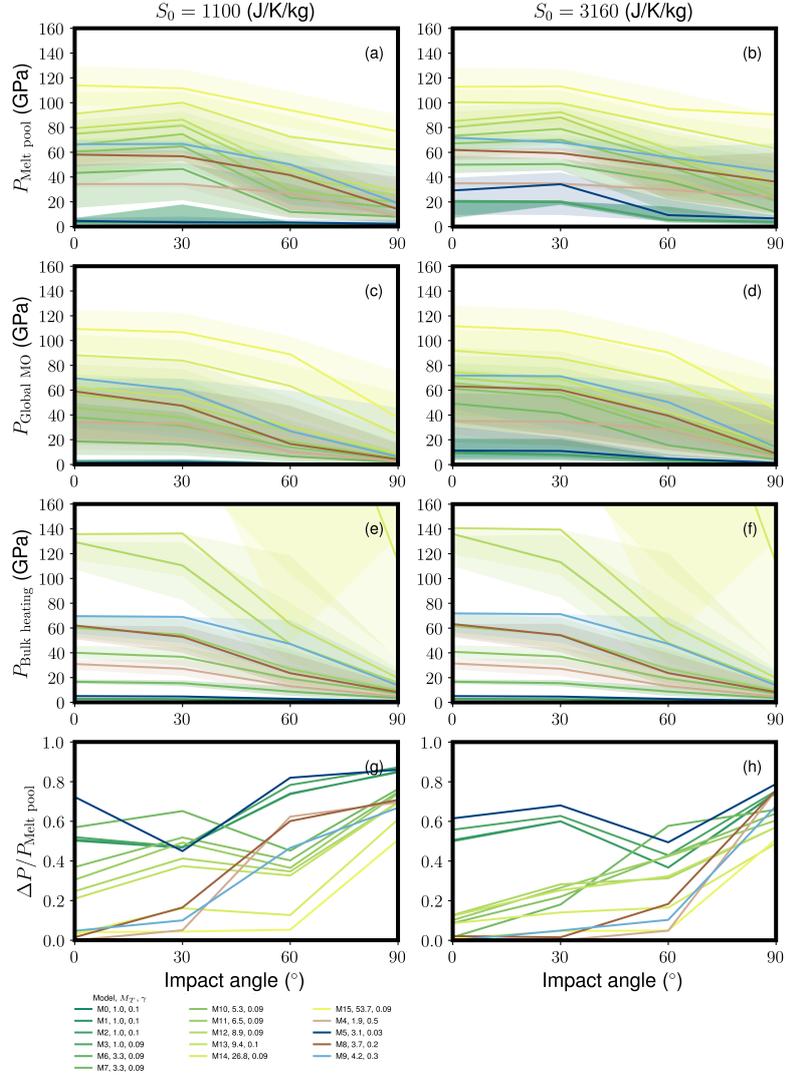}
  \end{center}
  \caption{The pressures at the bottom of melt for the $v_{\rm imp}= v_{\rm esc}$ cases. The left panels represent the initial mantle entropy of $S_0=1100$ J/K/kg, and the right panels represent $S_0=3160$ J/K/kg. (a, b) The pressure at the base of a melt pool, $P_{\rm Melt\ pool}$, (c, d) pressure at the base of a global magma ocean $P_{\rm Global\ MO}$, and (e, f) pressure of a melt whose volume is calculated from the bulk heating model, $P_{\rm Bulk\ heating}$ (see the main text), which does not depend on the initial temperature profile, and (g, h) the fractional difference in pressure between the melt pool and global magma ocean model ($\Delta P/P_{\rm Melt\ pool} = |P_{\rm Melt\ pool}-P_{\rm Global\ MO}|/P_{\rm Melt\ pool}$). The shaded regions in (a-f) show the error envelopes.}
  \label{fig:rp_vesc}
\end{figure*}

Figure \ref{fig:rp_vesc}a-f shows the pressures at the base of (1) a melt pool (panels a-b), (2) a global magma ocean (panels c-d), and (3) a global magma ocean whose melt volume is estimated by a bulk heating model (panels e-f) as discussed below for all $v_{\rm imp}=v_{\rm esc}$ cases. The left panels correspond to the $S_0=1100$ J/K/kg cases and the right panels correspond to the $S_0=3160$ J/K/kg cases. The shaded regions represent the error envelopes. 
In the melt pool model, we calculate its shape as discussed in Section \ref{sec:initT}. The pressure at the deepest portion of the melt pool is defined as $P_{\rm Melt\ pool}$.
In the global magma ocean model, the melt volume is the same as that of the melt pool, whereas the magma ocean depth is uniform (Figure \ref{fig:meltmodels}a) and the pressure at the base of the magma ocean is defined as $P_{\rm Global\  MO}$. In the bulk heating model, the melt mass fraction of the mantle, $f_{\rm melt}$, is described as
\begin{equation}
f_{\rm melt}= \frac{h(\theta)\Delta IE (\theta)}{M_{\rm mantle}(\theta) E_M},
\label{eq:meltfraction}
\end{equation}
where $E_M$ is the specific energy required for the material to melt after it experiences isentropic decompression down to 1 atm \citep{BjorkmanHolsapple1987}. 
Criteria of this type have been widely used in previous studies (e.g., \citealt{Abramovetal2012, PierazzoMelosh2000}). $P_{\rm Bulk\ heating}$ is defined as the pressure at the bottom of a magma ocean whose melt volume is $f_{\rm mantle}\frac{4 \pi}{3}(R_p^3-R_c^3)$, where $R_p$ is the radius and $R_c$ is the core radius of the post-impact body.
Figure \ref{fig:rp_vesc}g and h represent the fractional difference in pressure between the melt pool model and global magma ocean model, $\Delta P/P_{\rm Melt\ pool}$, where $\Delta P = |P_{\rm Melt\ pool}-P_{\rm Global\ MO}|$. The error envelopes are excluded in f-g panels.

As shown in the figure, $P_{\rm Melt\ pool}$ is usually greater than $P_{\rm Global\ MO}$, unless the entire mantle is molten. Higher $S_0$ causes forming deeper magma oceans as also seen in Figure \ref{fig:initialT}. Comparison between the bulk heating model and the other models is not straightforward because the depth of a magma ocean critically depends on the value of $E_M$, which is not well constrained. The fractional difference between $P_{\rm Melt\ pool}$ and $P_{\rm Global\ MO}$ can range from 0 to $\approx 0.8$ (panels g and h). The difference tends to be large at $\theta=90^\circ$, where the melt volume is smaller than those at the other angles and the geometry of a melt affects the pressure at the base of the melt region.

%

\section{Discussion}
\label{discussion}
\subsection{Python script for the melt scaling laws on GitHub}
\label{sec:GitHubModel}
We provide a Python script for this model on GitHub. In this script, $\Delta IE (\theta)$ is computed based on the fitting model we develop (Equation \ref{eq:IE}) as a function of the impact angle $\theta$, total mass $M_T$, impactor-to-total mass ratio $\gamma$, and impact velocity $v_{\rm imp}$. The options for the impact angle $\theta$ are $0^\circ, 30^\circ, 60^\circ$, and $90^\circ$, but we intend to include $45^\circ$ in the near future. Users can choose one of the options which is closest to the desired impact angle. Alternatively, users can interpolate the impact angles, but such interpolation has not been tested in this work.
Impact-induced heating is modeled using Equation \ref{eq:dU}. The initial thermal state of the mantle can be selected from the following options; (a) $S_0=1100$ J/K/kg, and (b) $S_0=3160$ J/K/kg, but in the future we are planning to add more options. So far we calculate the shape of a magma ocean as well as the pressure at the base of the melt using the melt criterion from \cite{Rubieetal2015}, but users can easily modify this criterion if necessary.

Our simulations do not include cases at $v_{\rm esc} < v_{\rm imp} < 1.1 v_{\rm esc}$, which is a common impact velocity range at the end of the planetary accretion phase (see Figure \ref{fig:nbody}). To take this into account, in the Python script, we use $v_{\rm cr}$, which is the critical velocity above which the target and impactor do not merge, as derived in \cite{Gendaetal2012}, 
\begin{equation}
\frac{v_{\rm cr}}{v_{\rm esc}}  = d_1 \Gamma^2 \Theta ^{d_5} + d_2 \Gamma + d_3 \Theta ^{d_5} + d_4,   
\label{eq:genda}
\end{equation}
where $\Gamma= (M_t-M_i)/M_T$ and $\Theta=1-\sin \theta$, $d_1=2.43, d_2=-0.0408, d_3 = 1.86, d_4 = 1.08,  d_5=5/2$.
When $v_{\rm imp} \leq v_{\rm cr}$, the low velocity parameters ($v_{\rm imp}=v_{\rm esc}$) are used, whereas when $v_{\rm imp} >  v_{\rm cr}$, the high velocity parameters ($v_{\rm imp} \geq 1.1 v_{\rm esc}$) are used.

\subsection{Model simplifications}
\label{sec:simplifications}
Our SPH simulations take into account the latent heat of silicate vaporization, but the effect is not explicitly considered in our analytical melt model. 
In most of our simulations, the vapor mass fractions (VMFs) are small ($<0.1$), but in some energetic cases, VMFs are large (0.1-1.0). In these cases, it is likely that the majority of the post-impact mantle experiences melting, and therefore a detailed melt scaling law may not be necessary when vaporization matters (see Section \ref{sec:additional_output_parameters} for detailed discussion). 

In our scaling law, we assume that the initial temperature profile of a planet does not affect the internal energy gain. For this reason, the total internal energy after an impact is calculated by simply adding an initial internal energy and internal energy gain obtained from our model. This is probably a fine assumption given that small changes in the internal energy do not affect the overall structure of a planet (such as the mass-radius relationship).
We also ignore the heating dependence on the azimuth ($\phi$) for simplicity. This should be fine for $\theta = 0^\circ$ cases, but this is not the case for other impact angles. We find that the dependence on $\phi$ is relatively weak (see Figure \ref{fig:xy_xz}), but this should be addressed in future studies.


\subsection{Implications for elemental partitioning}
\label{sec:partitioning}
Our new melt scaling laws describe the geometry of impact-induced melt for the first time. We hope that this will be valuable for the community and will be used to reevaluate whether planetary accretion models \citep[e.g.,][]{Rubieetal2015} are consistent with elemental abundances in Earth and other planetary objects (see discussion in Section \ref{intro}).
Moreover, the melt pool model predicts a higher pressure than a global magma ocean does, which would lead to higher $P-T$ conditions, thus more light elements, such as Mg, Si, and O, may be present in the Earth's core, which are considered to be partitioned into the core under high pressure and temperature conditions (e.g., \citealt{Siebertetal2013, Fischeretal2015, ORourkeStevenson2016}). This could affect heat flux, magnetic field, and seismic observations even today (e.g., \citealt{Labrosse2015}).

This paper only considers the initial condition right after the melt forming impact, but it is important to consider its time evolution to calculate element equilibration processes. \cite{deVriesetal2016} consider the evolution of magma oceans that become shallower over time due to crystallization. They find that a small impactor, which does not generate a magma ocean by itself, can still contribute to the metal-silicate equilibration process if it falls into a pre-existing magma ocean. Its depth is controlled not only by the initial melt volume but also by the time between the impacts.

\subsection{Material strength and choice of EOS}
\label{sec:materialstrength}

We do not consider the effect of material strength in this work, which can affect the extent of melting and heat distribution  \citep[e.g.,][]{Quintanaetal2015, Golabeketal2018, Emsenhuberetal2018, KurosawaGenda2018}. 
Material strength is known to matter for relatively small impacts and low impact velocities \citep{BenzAsphaug1999, Martin2015}, where friction and plastic deformation cause additional melting \citep{KurosawaGenda2018}.
This effect becomes less important for large impacts where shock heating overwhelms heating due to friction and plastic deformation \citep{MeloshIvanov2018}.

Another factor is the extent of deformation. If the peak pressure due to an impact exceeds the elastic limit ($0.1-10$ GPa, \citealt[e.g.,][]{Jeanloz1980}), treating the material as a fluid is appropriate. This is especially true near the impact site. However, this may not be appropriate for places far from the impact site. For example, the antipodal heating at $\theta=0^\circ$ is due to focusing of shock waves as well as potential energy release after an extensive deformation (see \ref{sec:SPH_simulations}). This extent of deformation and resulting heating is an overestimate considering the rigidity of a planet ($h_{\rm e}/h_{\rm f}\sim 0.01 (R/1000\ {\rm [km]})^2$, where $h_{\rm e}$ is the elastic tidal response height and $h_{\rm f}$ is the fluid tidal response height of a homogeneous spherical body, and $R$ is the radius of the body, \citealt{StevensonTextbook}, D. J., online textbook). 

The choice of EOS also affects the outcome. The Tillotson EOS is not an appropriate choice because it does not adequately describe the thermodynamics of the system. The choice of input parameters for M-ANEOS can be important because it affects the extent of shock heating and vaporization \citep{Stewartetal2020}. We will further investigate its effect in a future study.

\subsection{Resolution}
\label{sec:resolution}
Figure \ref{fig:resolution3} shows internal energy gains calculated in SPH for models M0, M1, and M2, whose resolutions are $N=10^4, 5\times10^4$, and $N=10^5$, where $N$ is the number of SPH particles. This shows that the $N=5\times10^4$ case is very similar to the $N=10^5$ case, but the $N=10^4$ case does not capture the details of the heat distribution very well. Based on these results, we mostly use the models with higher resolution than $N=10^4$.

\section{Conclusions}
\label{conclusions}
We develop mantle melt scaling laws as a function of the impact angle, impact velocity, total mass, and impactor-to-total mass ratio based on more than 100 SPH simulations. Our scaling laws include an analytical expression for the spatial heat distribution as a function of Legendre polynomials. Our scaling laws reproduce the heat distribution within a mantle computed by SPH simulations relatively well. We also find that the pressure difference at the base of a global (radially homogeneous) magma ocean, often used in literature, and melt pool, a spatially confined melt, can reach up to $\approx$80 \%.
This can have a significant impact on interpreting metal-silicate equilibration processes and therefore it would be important to revisit chemical evolution models of planetary mantle and core during the planet accretion phase. The scaling laws are publicly available via GitHub.

\section*{Acknowledgements}
\label{acknowledgement}

We first thank Jay Melosh, who passed away during the revision process, for his leadership in the impact community. We thank our two anonymous reviewers for helpful comments that significantly improved this manuscript. 
We also thank Alessandro Morbidelli, Francis Nimmo, and St\'{e}phane Labrosse for helpful discussions. We acknowledge support from the Carnegie DTM fellowship, the National Aeronautics and Space Administration under Grant No. 80NSSC19K0514.
Partial funding for this research was provided by the Center for Matter at Atomic Pressures (CMAP), a National Science Foundation (NSF) Physics Frontier Center, under Award PHY-2020249. Any opinions, findings, conclusions or recommendations expressed in this material are those of the authors and do not necessarily reflect those of the National Science Foundation. 
The project was supported in part by the Alfred P. Sloan Foundation under grant G202114194. This project is also supported by the European Research Council (ERC) Advanced Grant ``ACCRETE'' (Contract No. 290568) and by the Deutsche Forschungsgemeinschaft (DFG, German Research Foundation) Priority Programme SPP1833 “Building a Habitable Earth” (RU 1323/10-1). K.W. and  L.M. are funded by the DFG grant SFB-TRR 170 (C4), TRR-170 Pub. No. 75. C.B. appreciates support by DFG project 398488521. The calculations were partly carried out on the GRAPE system at the Center for Computational Astrophysics, National Astronomical Observatory of Japan. 
Figures \ref{fig:meltmodels} and \ref{fig:angledef} are created by Michael Franchot. The perceptually-uniform color maps used in Figures \ref{fig:Heatmodel}-\ref{fig:initialT} as well as Figures \ref{fig:xy_xz}, \ref{fig:resolution3}, and \ref{fig:M4d} were provided by Fabio Crameri \citep{Crameri2018}.


%

\bibliographystyle{model2-names}
\bibliography{main.bbl}

\newpage
\setcounter{page}{1}
\setcounter{section}{0}
\renewcommand{\thesection}{S.\arabic{section}}
\setcounter{figure}{0} 
\renewcommand{\thefigure}{S.\arabic{figure}}
\setcounter{table}{0} 
\renewcommand{\thetable}{S.\arabic{table}}
\setcounter{equation}{0} 
\renewcommand{\theequation}{S.\arabic{equation}}
\section*{Supplementary Materials}
\begin{center}
\begin{table*}[ht]
\tiny
\scalebox{1.0}{
\hfill{}
\begin{tabular}{c c c c c c c c c c c c c }
\hline
Model & ID &  $M_T$ & $\gamma$ & $\theta(^\circ)$ & $\frac{v_{\rm imp}}{v_{\rm esc}}$  & $v_{\rm esc}$ (m/s) & $dE$ (J) & $\frac{dE_{\rm mantle}}{dE}$ & $\frac{M_{\rm mantle}}{f_{\rm mantle} M_T}$ &  $MF_{\rm A}$  &  $\sigma^\prime$ & $N$\\
\hline
M0 & 1 & 1.0 & 0.1 & 0 & 1.0 & 4246 & 7.054e+29 & 0.754 & 0.998 & 0.639 & 0.8881 & 100000\\
M0 & 2 & 1.0 & 0.1 & 30 & 1.0 & 4246 & 6.478e+29 & 0.825 & 0.994 & 0.538 & 0.8982 & 100000\\
M0 & 3 & 1.0 & 0.1 & 60 & 1.0 & 4246 & 4.59e+29 & 0.855 & 0.975 & 0.273 & 0.8374 & 100000\\
M0 & 4 & 1.0 & 0.1 & 90 & 1.0 & 4246 & 1.149e+29 & 0.861 & 0.914 & 0.087 & 0.559 & 100000\\
M1 & 5 & 1.0 & 0.1 & 0 & 1.0 & 4246 & 6.983e+29 & 0.729 & 0.998 & 0.639 & 0.963 & 10000\\
M1 & 6 & 1.0 & 0.1 & 30 & 1.0 & 4246 & 7.162e+29 & 0.799 & 0.997 & 0.538 & 0.9404 & 10000\\
M1 & 7 & 1.0 & 0.1 & 60 & 1.0 & 4246 & 4.844e+29 & 0.822 & 0.978 & 0.273 & 0.9006 & 10000\\
M1 & 8 & 1.0 & 0.1 & 90 & 1.0 & 4246 & 3.345e+29 & 0.834 & 0.952 & 0.087 & 0.864 & 10000\\
M2 & 9 & 1.0 & 0.1 & 0 & 1.0 & 4246 & 7.229e+29 & 0.752 & 0.998 & 0.639 & 0.9269 & 50000\\
M2 & 10 & 1.0 & 0.1 & 30 & 1.0 & 4246 & 6.71e+29 & 0.809 & 0.996 & 0.538 & 0.9296 & 50000\\
M2 & 11 & 1.0 & 0.1 & 60 & 1.0 & 4246 & 5.129e+29 & 0.83 & 0.985 & 0.273 & 0.7754 & 50000\\
M2 & 12 & 1.0 & 0.1 & 90 & 1.0 & 4246 & 1.7e+29 & 0.777 & 0.914 & 0.087 & 0.5488 & 50000\\
M3 & 13 & 1.03 & 0.091 & 0 & 1.0 & 4401 & 6.7e+29 & 0.731 & 0.999 & 0.588 & 0.96 & 16500\\
M3 & 14 & 1.03 & 0.091 & 30 & 1.0 & 4401 & 6.35e+29 & 0.805 & 0.997 & 0.509 & 0.9492 & 16500\\
M3 & 15 & 1.03 & 0.091 & 60 & 1.0 & 4401 & 4.342e+29 & 0.84 & 0.977 & 0.251 & 0.9458 & 16500\\
M3 & 16 & 1.03 & 0.091 & 90 & 1.0 & 4401 & 1.283e+29 & 0.866 & 0.923 & 0.079 & 0.6542 & 16500\\
M4 & 17 & 1.88 & 0.5 & 0 & 1.0 & 5061 & 4.778e+30 & 0.725 & 0.992 & 1.000 & 1.665 & 30000\\
M4 & 18 & 1.88 & 0.5 & 30 & 1.0 & 5061 & 3.739e+30 & 0.775 & 0.979 & 1.000 & 2.082 & 30000\\
M4 & 19 & 1.88 & 0.5 & 60 & 1.0 & 5061 & 2.114e+30 & 0.813 & 0.937 & 0.901 & 1.69 & 30000\\
M4 & 20 & 1.88 & 0.5 & 90 & 1.0 & 5061 & 1.469e+30 & 0.777 & 0.92 & 0.289 & 1.455 & 30000\\
M5 & 21 & 3.06 & 0.032 & 0 & 1.0 & 6622 & 1.671e+30 & 0.804 & 1.0 & 0.356 & 1.065 & 31000\\
M5 & 22 & 3.06 & 0.032 & 30 & 1.0 & 6622 & 1.458e+30 & 0.852 & 0.998 & 0.351 & 1.021 & 31000\\
M5 & 23 & 3.06 & 0.032 & 60 & 1.0 & 6622 & 7.423e+29 & 0.859 & 0.982 & 0.178 & 0.6705 & 31000\\
M5 & 24 & 3.06 & 0.032 & 90 & 1.0 & 6622 & 1.635e+29 & 0.859 & 0.97 & 0.055 & 0.5632 & 31000\\
M6 & 25 & 3.25 & 0.091 & 0 & 1.0 & 6473 & 4.697e+30 & 0.726 & 0.999 & 0.922 & 1.333 & 33000\\
M6 & 26 & 3.25 & 0.091 & 30 & 1.0 & 6473 & 4.115e+30 & 0.814 & 0.996 & 0.856 & 1.201 & 33000\\
M6 & 27 & 3.25 & 0.091 & 60 & 1.0 & 6473 & 2.874e+30 & 0.834 & 0.975 & 0.453 & 0.9588 & 33000\\
M6 & 28 & 3.25 & 0.091 & 90 & 1.0 & 6473 & 6.881e+29 & 0.855 & 0.921 & 0.146 & 0.6311 & 33000\\
M7 & 29 & 3.25 & 0.091 & 0 & 1.0 & 6527 & 4.558e+30 & 0.755 & 0.999 & 0.922 & 1.294 & 33000\\
M7 & 30 & 3.25 & 0.091 & 30 & 1.0 & 6527 & 3.782e+30 & 0.822 & 0.993 & 0.856 & 1.352 & 33000\\
M7 & 31 & 3.25 & 0.091 & 60 & 1.0 & 6527 & 2.917e+30 & 0.858 & 0.976 & 0.453 & 0.8844 & 33000\\
M7 & 32 & 3.25 & 0.091 & 90 & 1.0 & 6527 & 4.55e+29 & 1.019 & 0.92 & 0.146 & 0.7225 & 33000\\
M8 & 33 & 3.7 & 0.2 & 0 & 1.0 & 6546 & 1.114e+31 & 0.753 & 0.998 & 0.998 & 1.846 & 37500\\
M8 & 34 & 3.7 & 0.2 & 30 & 1.0 & 6546 & 8.654e+30 & 0.817 & 0.992 & 0.962 & 1.619 & 37500\\
M8 & 35 & 3.7 & 0.2 & 60 & 1.0 & 6546 & 8.354e+30 & 0.736 & 0.96 & 0.799 & 1.507 & 37500\\
M8 & 36 & 3.7 & 0.2 & 90 & 1.0 & 6546 & 4.688e+30 & 0.807 & 0.915 & 0.270 & 1.536 & 37500\\
M9 & 37 & 4.23 & 0.301 & 0 & 1.0 & 6757 & 1.681e+31 & 0.772 & 0.993 & 1.000 & 2.186 & 42900\\
M9 & 38 & 4.23 & 0.301 & 30 & 1.0 & 6757 & 1.392e+31 & 0.793 & 0.981 & 0.985 & 2.207 & 42900\\
M9 & 39 & 4.23 & 0.301 & 60 & 1.0 & 6757 & 8.701e+30 & 0.785 & 0.942 & 0.861 & 1.844 & 42900\\
M9 & 40 & 4.23 & 0.301 & 90 & 1.0 & 6757 & 6.898e+30 & 0.775 & 0.921 & 0.372 & 1.836 & 42900\\
\hline
\end{tabular}}
\hfill{}
\caption{List of parameters at $v_{\rm imp}=v_{\rm esc}$.
Model represents a set of impact simulations that have the same parameters except impact angles ($0^\circ, 30^\circ, 60^\circ,$ and $90^\circ$). ID represents each SPH simulation. $M_T$ is the total mass normalized by the Martian mass, $\gamma$ is the impactor-to-total mass ratio, $\theta$ is the impact angle in degrees ($0^\circ$ is a head-on collision). $v_{\rm imp}$ is the impact velocity, and $v_{\rm esc}$ is the mutual escape velocity in m/s.
$dE$ is the impact-induced energy in J, and $dE_{\rm mantle}/dE$ is the fraction of the energy that goes into the mantle, and $M_{\rm mantle}$ is the post-impact mantle mass, $f_{\rm mantle}$ is the mantle mass fraction (0.7), $MF_{\rm A}$ is the calculated melt mass fraction of the mantle with the initial mantle entropy of 3160 J/K/kg. $\sigma^\prime$ is the error between the model and SPH. $N$ is the number of SPH particles.}
\label{tb:list1}
\end{table*}
\end{center}

\begin{center}
\begin{table*}[ht]
\tiny
\scalebox{1.0}{
\hfill{}
\begin{tabular}{c c c c c c c c c c c c c }
\hline
Model & ID &  $M_T$ & $\gamma$ & $\theta(^\circ)$ & $\frac{v_{\rm imp}}{v_{\rm esc}}$  & $v_{\rm esc}$ & dE (J) & $\frac{dE_{\rm mantle}}{dE}$ & $\frac{M_{\rm mantle}}{f_{\rm mantle} M_T}$  & $MF_{\rm A}$  & $\sigma^\prime$ & $N$\\
\hline
M10 & 41 & 5.34 & 0.091 & 0 & 1.0 & 7698 & 9.691e+30 & 0.739 & 0.999 & 0.848 & 1.85 & 33000\\
M10 & 42 & 5.34 & 0.091 & 30 & 1.0 & 7698 & 9.23e+30 & 0.812 & 0.995 & 0.799 & 1.469 & 33000\\
M10 & 43 & 5.34 & 0.091 & 60 & 1.0 & 7698 & 2.545e+30 & 0.869 & 0.929 & 0.528 & 1.354 & 33000\\
M10 & 44 & 5.34 & 0.091 & 90 & 1.0 & 7698 & 1.395e+30 & 0.903 & 0.921 & 0.185 & 0.8959 & 33000\\
M11 & 45 & 6.54 & 0.091 & 0 & 1.0 & 8271 & 1.506e+31 & 0.724 & 0.999 & 0.803 & 1.816 & 33000\\
M11 & 46 & 6.54 & 0.091 & 30 & 1.0 & 8271 & 1.287e+31 & 0.814 & 0.995 & 0.759 & 1.446 & 33000\\
M11 & 47 & 6.54 & 0.091 & 60 & 1.0 & 8271 & 3.599e+30 & 0.865 & 0.929 & 0.520 & 1.483 & 33000\\
M11 & 48 & 6.54 & 0.091 & 90 & 1.0 & 8271 & 1.897e+30 & 0.913 & 0.921 & 0.199 & 0.8389 & 33000\\
M12 & 49 & 8.94 & 0.091 & 0 & 1.0 & 9241 & 3.068e+31 & 0.731 & 0.999 & 0.721 & 1.87 & 33000\\
M12 & 50 & 8.94 & 0.091 & 30 & 1.0 & 9241 & 2.993e+31 & 0.811 & 0.987 & 0.684 & 1.639 & 33000\\
M12 & 51 & 8.94 & 0.091 & 60 & 1.0 & 9241 & 7.589e+30 & 0.893 & 0.925 & 0.504 & 1.54 & 33000\\
M12 & 52 & 8.94 & 0.091 & 90 & 1.0 & 9241 & 4.332e+30 & 0.929 & 0.921 & 0.221 & 0.9747 & 33000\\
M13 & 53 & 9.43 & 0.104 & 0 & 1.0 & 9217 & 3.599e+31 & 0.692 & 1.0 & 0.734 & 2.167 & 33482\\
M13 & 54 & 9.43 & 0.104 & 30 & 1.0 & 9217 & 3.392e+31 & 0.769 & 0.997 & 0.703 & 1.788 & 33482\\
M13 & 55 & 9.43 & 0.104 & 60 & 1.0 & 9217 & 2.402e+31 & 0.755 & 0.972 & 0.524 & 1.534 & 33482\\
M13 & 56 & 9.43 & 0.104 & 90 & 1.0 & 9217 & 4.758e+30 & 0.865 & 0.913 & 0.241 & 1.191 & 33482\\
M14 & 57 & 26.84 & 0.091 & 0 & 1.0 & 13600 & 1.73e+32 & 0.766 & 1.0 & 0.467 & 2.877 & 11000\\
M14 & 58 & 26.84 & 0.091 & 30 & 1.0 & 13600 & 1.56e+32 & 0.82 & 0.996 & 0.454 & 2.487 & 11000\\
M14 & 59 & 26.84 & 0.091 & 60 & 1.0 & 13600 & 5.664e+31 & 0.907 & 0.946 & 0.391 & 2.086 & 11000\\
M14 & 60 & 26.84 & 0.091 & 90 & 1.0 & 13600 & 2.472e+31 & 0.919 & 0.923 & 0.222 & 1.363 & 11000\\
M15 & 61 & 53.66 & 0.091 & 0 & 1.0 & 17520 & 5.865e+32 & 0.766 & 1.0 & 0.347 & 3.562 & 11000\\
M15 & 62 & 53.66 & 0.091 & 30 & 1.0 & 17520 & 5.206e+32 & 0.816 & 0.996 & 0.339 & 2.807 & 11000\\
M15 & 63 & 53.66 & 0.091 & 60 & 1.0 & 17520 & 1.709e+32 & 0.907 & 0.942 & 0.305 & 2.938 & 11000\\
M15 & 64 & 53.66 & 0.091 & 90 & 1.0 & 17520 & 7.895e+31 & 0.917 & 0.922 & 0.178 & 1.843 & 11000\\
\hline
\end{tabular}}
\hfill{}
\caption{Continuation of Table \ref{tb:list1}.}
\label{tb:list2}
\end{table*}
\end{center}

\begin{center}
\begin{table*}[ht]
\tiny
\scalebox{1.0}{
\hfill{}
\begin{tabular}{c c c c c c c c c c c c c }
\hline
Model & ID &  $M_T$ & $\gamma$ & $\theta(^\circ)$ & $\frac{v_{\rm imp}}{v_{\rm esc}}$  & $v_{\rm esc}$ & dE (J) & $\frac{dE_{\rm mantle}}{dE}$ & $\frac{M_{\rm mantle}}{f_{\rm mantle} M_T}$ & $MF_{\rm A}$ & $\sigma^\prime$ & $N$\\
\hline
M16 & 65 & 1.88 & 0.5 & 0 & 1.3 & 5061 & 5.624e+30 & 0.75 & 0.937 & 1.000 & 1.659 & 30000\\
M16 & 66 & 1.88 & 0.5 & 30 & 1.3 & 5061 & 4.744e+30 & 0.818 & 0.982 & 1.000 & 2.736 & 30000\\
M16 & 67 & 1.88 & 0.5 & 60 & 1.3 & 5061 & 4.551e+29 & 0.798 & 0.498 & 0.608 & 1.438 & 30000\\
M16 & 68 & 1.88 & 0.5 & 90 & 1.3 & 5061 & 1.855e+29 & 0.65 & 0.5 & 0.127 & 0.8505 & 30000\\
M17 & 69 & 3.25 & 0.091 & 0 & 1.1 & 6473 & 5.012e+30 & 0.76 & 0.999 & 0.950 & 1.118 & 33000\\
M17 & 70 & 3.25 & 0.091 & 30 & 1.1 & 6473 & 4.47e+30 & 0.823 & 0.993 & 0.886 & 1.072 & 33000\\
M17 & 71 & 3.25 & 0.091 & 60 & 1.1 & 6473 & 1.113e+30 & 0.863 & 0.926 & 0.095 & 0.6728 & 33000\\
M17 & 72 & 3.25 & 0.091 & 90 & 1.1 & 6473 & 3.067e+29 & 0.758 & 0.912 & 0.011 & 0.4547 & 33000\\
M18 & 73 & 3.25 & 0.091 & 0 & 1.2 & 6473 & 5.724e+30 & 0.77 & 0.997 & 0.974 & 1.065 & 33000\\
M18 & 74 & 3.25 & 0.091 & 30 & 1.2 & 6473 & 4.802e+30 & 0.826 & 0.987 & 0.908 & 0.9694 & 33000\\
M18 & 75 & 3.25 & 0.091 & 60 & 1.2 & 6473 & 1.023e+30 & 0.87 & 0.922 & 0.125 & 0.7102 & 33000\\
M18 & 76 & 3.25 & 0.091 & 90 & 1.2 & 6473 & 2.657e+29 & 0.753 & 0.911 & 0.016 & 0.4027 & 33000\\
M19 & 77 & 3.25 & 0.091 & 0 & 1.3 & 6473 & 6.224e+30 & 0.796 & 0.996 & 0.994 & 1.266 & 33000\\
M19 & 78 & 3.25 & 0.091 & 30 & 1.3 & 6473 & 5.108e+30 & 0.828 & 0.982 & 0.933 & 0.9734 & 33000\\
M19 & 79 & 3.25 & 0.091 & 60 & 1.3 & 6473 & 1.298e+30 & 0.81 & 0.919 & 0.161 & 0.5949 & 33000\\
M19 & 80 & 3.25 & 0.091 & 90 & 1.3 & 6473 & 4.493e+29 & 0.646 & 0.911 & 0.023 & 0.4536 & 33000\\
M20 & 81 & 3.25 & 0.091 & 0 & 1.4 & 6473 & 6.987e+30 & 0.811 & 0.996 & 0.999 & 1.219 & 33000\\
M20 & 82 & 3.25 & 0.091 & 30 & 1.4 & 6473 & 5.274e+30 & 0.831 & 0.974 & 0.960 & 1.005 & 33000\\
M20 & 83 & 3.25 & 0.091 & 60 & 1.4 & 6473 & 9.941e+29 & 0.876 & 0.917 & 0.187 & 0.6842 & 33000\\
M20 & 84 & 3.25 & 0.091 & 90 & 1.4 & 6473 & 1.672e+29 & 0.777 & 0.91 & 0.027 & 0.4518 & 33000\\
M21 & 85 & 3.25 & 0.091 & 0 & 1.5 & 6473 & 8.071e+30 & 0.809 & 0.993 & 1.000 & 1.094 & 33000\\
M21 & 86 & 3.25 & 0.091 & 30 & 1.5 & 6473 & 5.972e+30 & 0.836 & 0.972 & 0.980 & 1.081 & 33000\\
M21 & 87 & 3.25 & 0.091 & 60 & 1.5 & 6473 & 1.006e+30 & 0.873 & 0.916 & 0.213 & 0.5735 & 33000\\
M21 & 88 & 3.25 & 0.091 & 90 & 1.5 & 6473 & 1.548e+29 & 0.784 & 0.91 & 0.036 & 0.5047 & 33000\\
M22 & 89 & 3.25 & 0.091 & 0 & 2.0 & 6473 & 1.232e+31 & 0.838 & 0.977 & 1.000 & 1.757 & 33000\\
M22 & 90 & 3.25 & 0.091 & 30 & 2.0 & 6473 & 6.664e+30 & 0.848 & 0.927 & 0.999 & 1.742 & 33000\\
M22 & 91 & 3.25 & 0.091 & 60 & 2.0 & 6473 & 9.251e+29 & 0.909 & 0.912 & 0.342 & 1.004 & 33000\\
M22 & 92 & 3.25 & 0.091 & 90 & 2.0 & 6473 & 1.138e+29 & 0.817 & 0.909 & 0.070 & 0.7277 & 33000\\
M23 & 93 & 4.23 & 0.301 & 0 & 1.1 & 6757 & 1.623e+31 & 0.783 & 0.981 & 1.000 & 2.381 & 42900\\
M23 & 94 & 4.23 & 0.301 & 30 & 1.1 & 6757 & 1.353e+31 & 0.801 & 0.968 & 0.997 & 1.466 & 42900\\
M23 & 95 & 4.23 & 0.301 & 60 & 1.1 & 6757 & 1.958e+30 & 0.888 & 0.712 & 0.464 & 1.043 & 42900\\
M23 & 96 & 4.23 & 0.301 & 90 & 1.1 & 6757 & 8.952e+29 & 0.834 & 0.703 & 0.093 & 0.6603 & 42900\\
M24 & 97 & 4.23 & 0.301 & 0 & 1.2 & 6757 & 1.62e+31 & 0.775 & 0.962 & 1.000 & 2.502 & 42900\\
M24 & 98 & 4.23 & 0.301 & 30 & 1.2 & 6757 & 1.341e+31 & 0.804 & 0.94 & 0.998 & 1.732 & 42900\\
M24 & 99 & 4.23 & 0.301 & 60 & 1.2 & 6757 & 1.556e+30 & 0.942 & 0.708 & 0.514 & 1.29 & 42900\\
M24 & 100 & 4.23 & 0.301 & 90 & 1.2 & 6757 & 5.99e+29 & 0.888 & 0.701 & 0.107 & 0.8105 & 42900\\
M25 & 101 & 4.23 & 0.301 & 0 & 1.3 & 6757 & 1.949e+31 & 0.776 & 0.957 & 1.000 & 1.895 & 42900\\
M25 & 102 & 4.23 & 0.301 & 30 & 1.3 & 6757 & 1.543e+31 & 0.811 & 0.937 & 0.999 & 1.945 & 42900\\
M25 & 103 & 4.23 & 0.301 & 60 & 1.3 & 6757 & 1.48e+30 & 0.958 & 0.706 & 0.577 & 1.457 & 42900\\
M25 & 104 & 4.23 & 0.301 & 90 & 1.3 & 6757 & 6.645e+29 & 0.826 & 0.701 & 0.118 & 0.9254 & 42900\\
M26 & 105 & 4.23 & 0.301 & 0 & 1.6 & 6757 & 1.886e+31 & 0.772 & 0.86 & 1.000 & 2.536 & 42900\\
M26 & 106 & 4.23 & 0.301 & 30 & 1.6 & 6757 & 7.866e+30 & 0.818 & 0.691 & 1.000 & 2.829 & 42900\\
M26 & 107 & 4.23 & 0.301 & 60 & 1.6 & 6757 & 1.843e+30 & 0.925 & 0.702 & 0.785 & 1.778 & 42900\\
M26 & 108 & 4.23 & 0.301 & 90 & 1.6 & 6757 & 3.892e+29 & 0.996 & 0.7 & 0.153 & 1.185 & 42900\\
\hline
\end{tabular}}
\hfill{}
\caption{The list of parameters is the same as those in Table \ref{tb:list1}, but for $v_{\rm imp} \geq 1.1 v_{\rm esc}$. }
\label{tb:list3}
\end{table*}
\end{center}

\label{supplementary}
\section{Model descriptions}
\label{sec:modeldetail}
\begin{figure*}
  \begin{center}
    \includegraphics[scale=2.0]{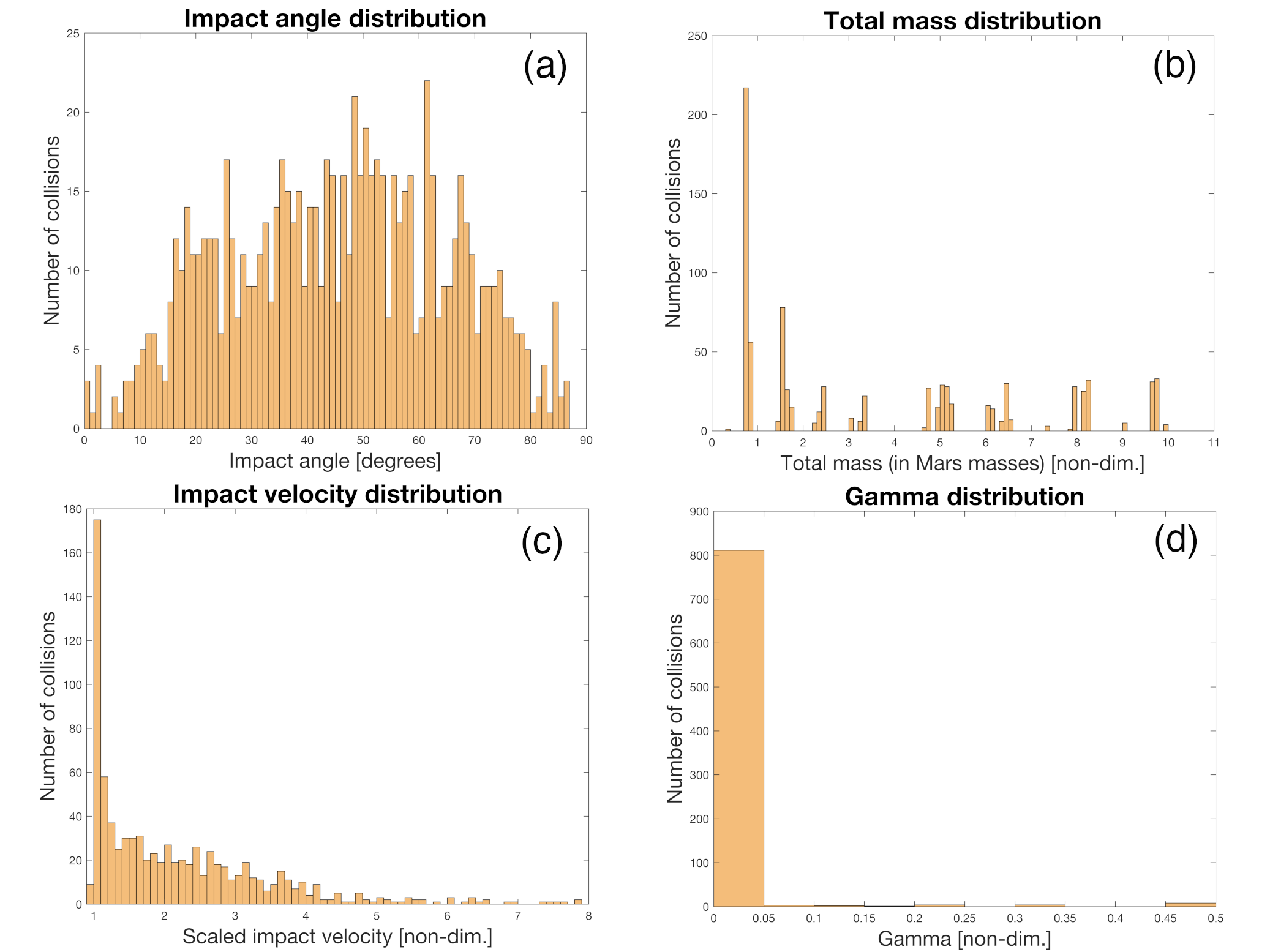}
  \end{center}
  \caption{Statistics of an impact simulation (run ``8:1-0.8-8'' from \citealt{Rubieetal2015}). The panels represent (a) impact angle distribution, (b) total mass distribution, (c) impact velocity distribution and (d) impactor-to-total mass ratio, $\gamma$.}
\label{fig:nbody}
\end{figure*}

In this section, we discuss model descriptions further.
Figure \ref{fig:nbody} shows the distribution of impact parameters taken from an $N-$body simulation (run ``8:1-0.8-8'', \citealt{Rubieetal2015}). The other simulations in \cite{Rubieetal2015} also have similar parameter distributions.
Our model reasonably covers these ranges (the typical ranges are $30^\circ<\theta<60^\circ$, $M_T<1-2 M_{\rm Mars}$, $v_{\rm imp} < 1.5 v_{\rm esc}$, and $\gamma < 0.05$), but high impact velocities ($>2 v_{\rm esc}$) have not been tested in our simulations. 

\begin{figure*}
  \begin{center}
    \includegraphics[scale=0.6]{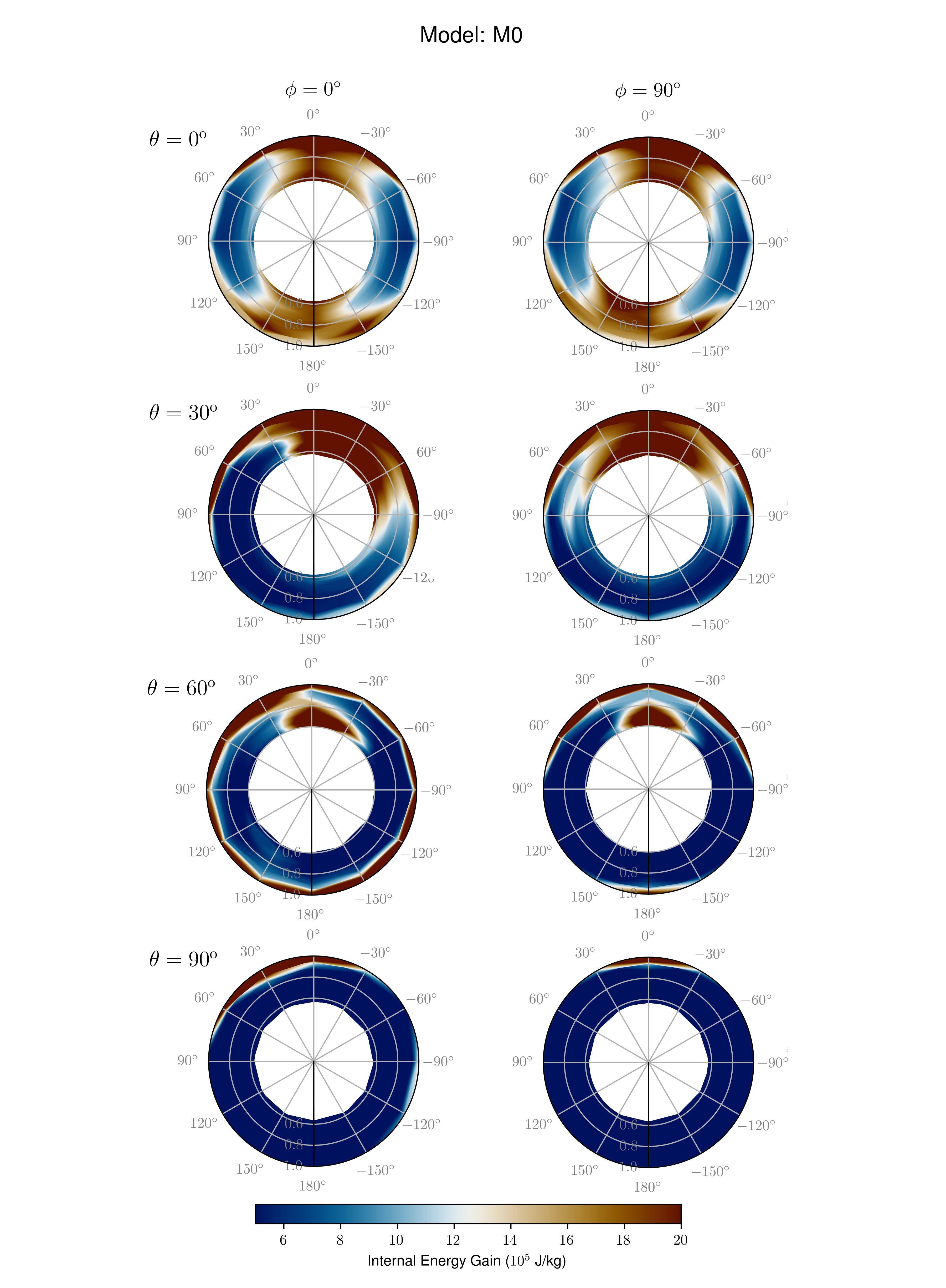}
  \end{center}
  \caption{The left panel shows the heat distribution at $\phi=0^\circ$ (the impact plane) and the right panel represents that at $\phi=90^\circ$ (the plane which is perpendicular to the impact plane) for model M0. Their heat distributions look reasonably similar, which suggests that our approximation that the heat distribution along the $\phi$ axis is symmetric is reasonable.}
  \label{fig:xy_xz}
\end{figure*}

Figure \ref{fig:xy_xz} shows the internal energy gain distribution for model M0 based on the SPH simulation. $\phi=0^\circ$ is the plane that is parallel to the impact velocity vector (physics should be axi-symmetric around the plane). $\phi=90^\circ$ is perpendicular to the plane $\phi=0^\circ$. The internal energy here is the averaged value within $0.2R_p$ from the plane, where $R_p$ is the planetary radius. In our model, we assume that the heat distribution does not depend on $\phi$ as discussed in Section \ref{model2}. This should be an accurate model for the $\theta=0^\circ$ cases, whereas this is an assumption for the $\theta>0^\circ$ cases. At $\theta=30^\circ$ and $90^\circ$, the dependence of the heat distribution on $\phi$ is relatively minor, but at $\theta=60^\circ$, the outer portion of the mantle at $\phi=0^\circ$ is more heated than that at $\phi=90^\circ$. This is because the impactor forms small fragments at this impact angle and the fragments nearly uniformly accrete onto the equator of the planet (see Figure \ref{fig:SPH}). This is not the case at $\phi=90^\circ$ and this is a limitation of our model.

\begin{figure*}
  \begin{center}
    \includegraphics[scale=0.6]{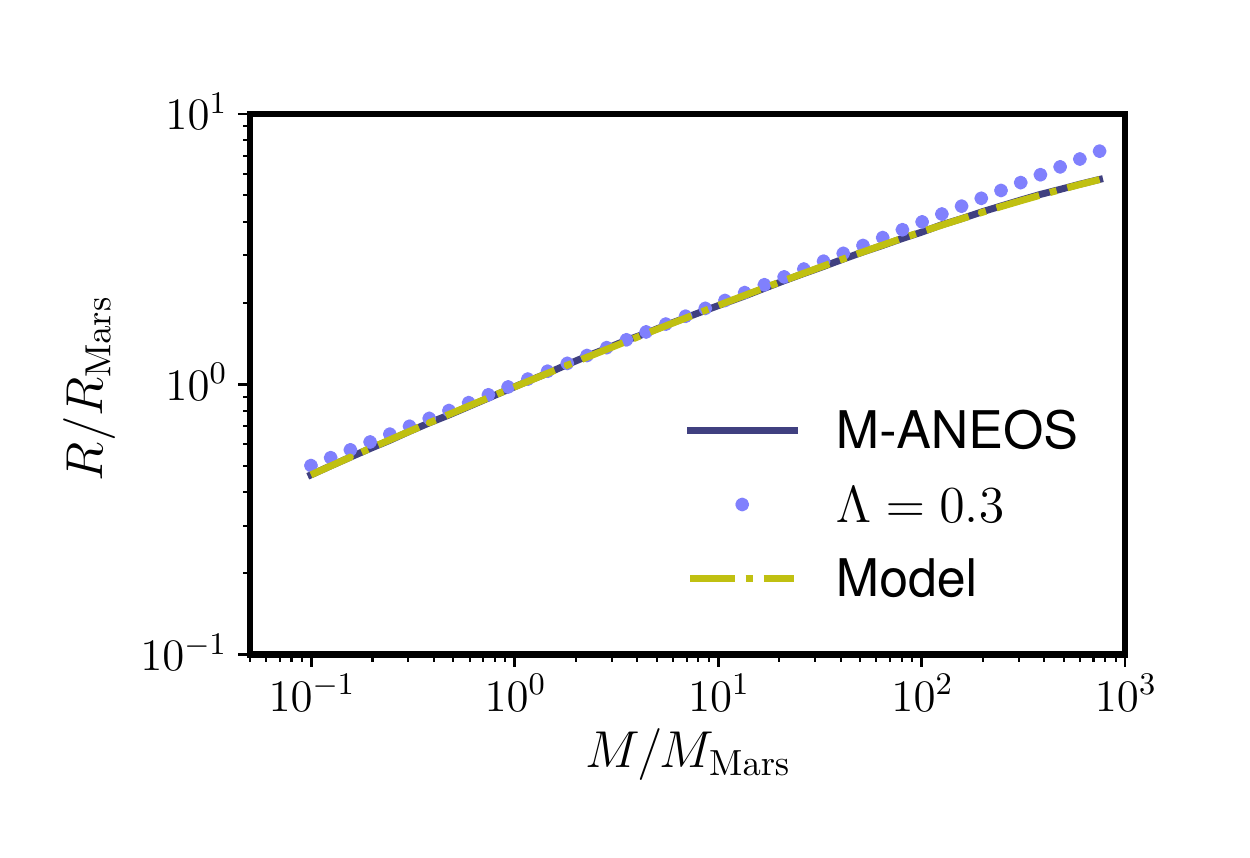}
  \end{center}
  \caption{Mass-radius relationship for planets with an adiabatic mantle and core. The core mass fraction is 0.3. The dark blue solid line represents our calculation using M-ANEOS and the dark yellow dashed line represents our fit to the calculation. The light blue dotted line represents $M\propto R^{0.3}$.}
\label{fig:mass_radius}
\end{figure*}

The mass-radius relationship is shown in Figure \ref{fig:mass_radius}. As discussed in Section \ref{model1}, the planetary radius, $R'$, of a planet whose mass is $M_t + M_i$ is described as 
$R'=R_0[(M_t + M_i)/M_0]^{\Lambda(M)}$, where $R_0=1.5717\times 10^6$ m and  $M_0=6.39\times10^{22}$ kg and $\Lambda(M)=\sum_{i=0}^{3}{b_i [\ln(M/M_0)]^i}$. The values of $b_i$ are listed in Table \ref{tb:parameterlist_MR}. 
We compute the mass-radius relationship of a planet by integrating the mass of a thin shell ($4 \pi \rho r^2 dr$) outwards with an initial guess of the central pressure until the pressure reaches zero. The value of the central pressure is modified until the calculated planetary mass reaches the target mass. The corresponding radius is also calculated during the process. The pressure-density relationship is determined along the isentrope based on M-ANEOS.
The entropy values for the mantle and core are 3160 J/K/kg and 1500 J/K/kg, respectively. We set the minimum mantle density to $3239$ kg/m$^3$.

\begin{figure*}
  \begin{center}
    \includegraphics[scale=0.6]{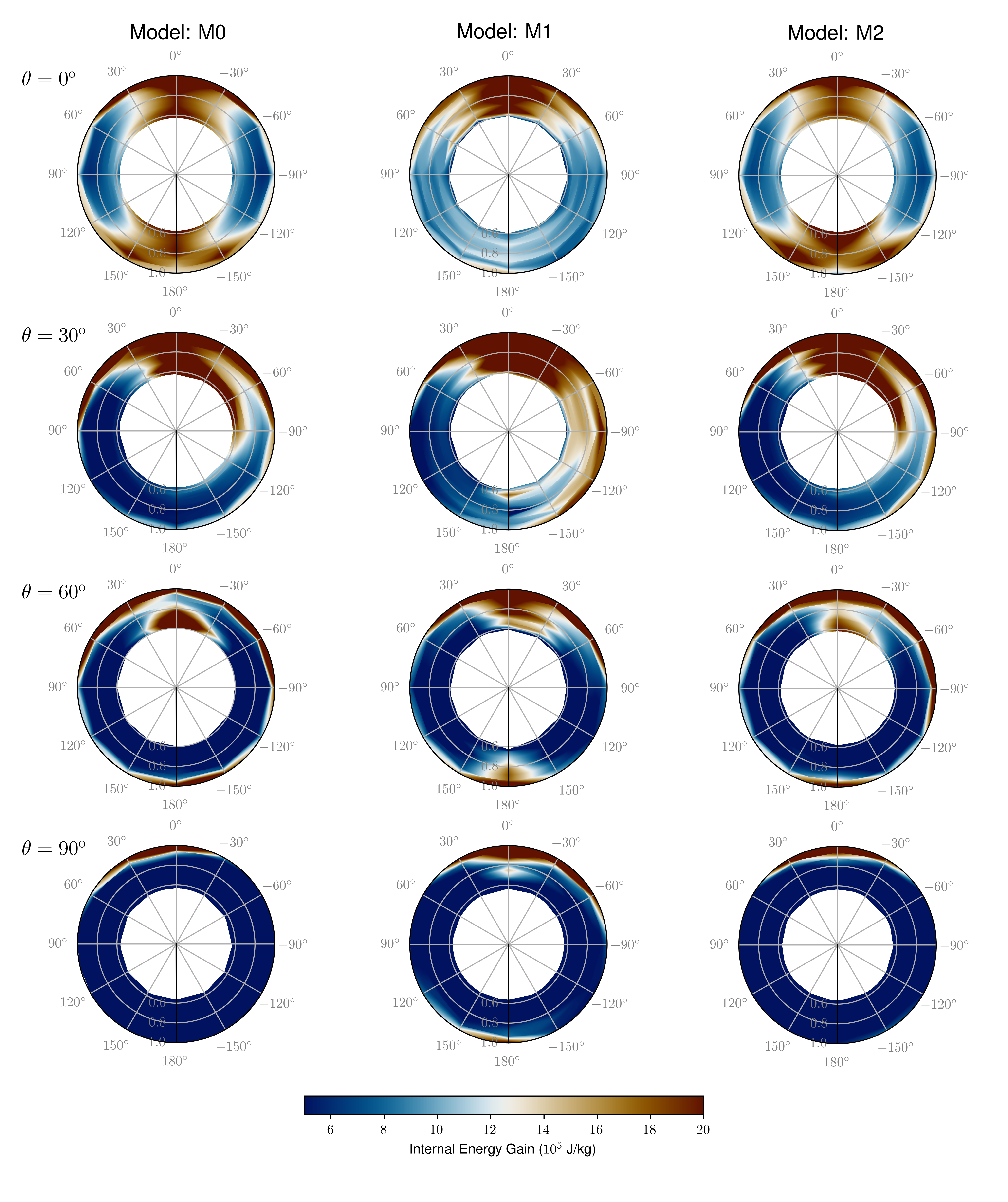}
  \end{center}
  \caption{Internal energy gain for Model M0  ($N=10^5$, Figure \ref{fig:M0d}), models M1 ($N=10^4$), and M2  ($N=5\times 10^4$). All the input parameters are the same. The overall heat distributions are similar between models M0 and M2, but they are different from that of model M1, which has the lowest resolution. }    
  \label{fig:resolution3}
\end{figure*}

Figure \ref{fig:resolution3} shows internal energies in $10^5$ J/kg for models M0, M1, and M2, whose resolutions are $N=10^4, N=5\times 10^4, 10^5$, respectively. The details are discussed in Section \ref{sec:resolution}. 

\begin{center}
\begin{table*}[ht]
\tiny
\scalebox{1.0}{
\hfill{}
\begin{tabular}{c c c c  }
\hline
$b_0$ & $b_1$ &  $b_2$ & $b_3$   \\
\hline
  \\
0.3412 & $-8.90\times 10^{-3}$ & $9.1442 \times 10^{-4}$  & $-7.4332 \times 10^{-5}$ \\
\hline
\end{tabular}}
\hfill{}
\caption{List of parameters to describe the planetary mass-radius relationship.}
\label{tb:parameterlist_MR}
\end{table*}
\end{center}

\begin{figure*}
  \begin{center}
    \includegraphics[scale=0.4]{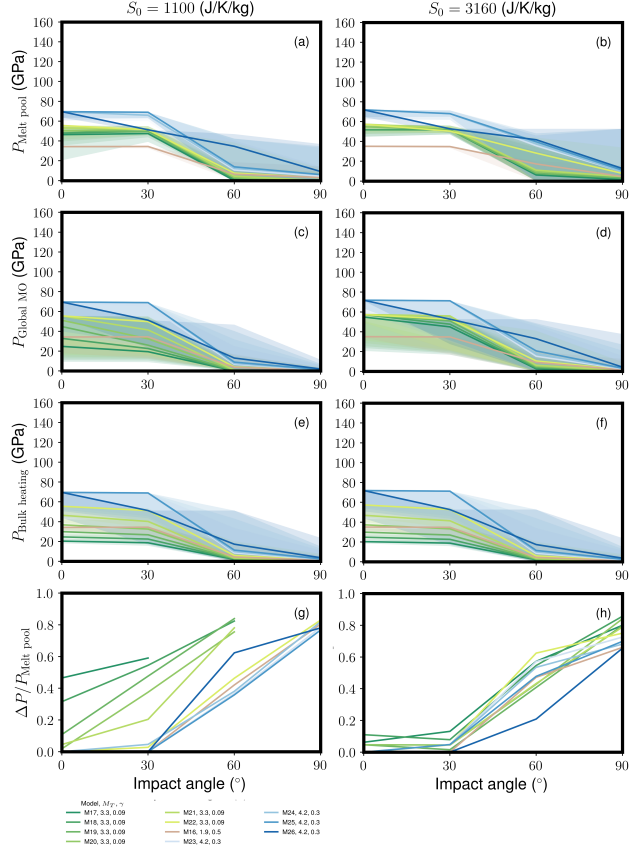}
  \end{center}
  \caption{Pressures at the base of (a, b) a melt pool, (c, d) a global magma ocean, (e,f) melt whose volume is calculated using the bulk heating model, and (g, h) the fractional difference in pressure between the melt pool and global magma ocean model ($\Delta P/P_{\rm Melt\ pool} = |P_{\rm Melt\ pool}-P_{\rm Global\ MO}|/P_{\rm Melt\ pool}$) for the $v_{\rm imp}\geq 1.1 v_{\rm esc}$ cases. All the parameters are the same as those in Figure \ref{fig:rp_vesc}. In panel g, some lines are missing because $P_{\rm Melt\ pool}=0$ (no melt). The errors are shown in shade in panels a-f.}
  \label{fig:abramov_vlarge}
\end{figure*}

\begin{figure*}
  \begin{center}
    \includegraphics[scale=0.05]{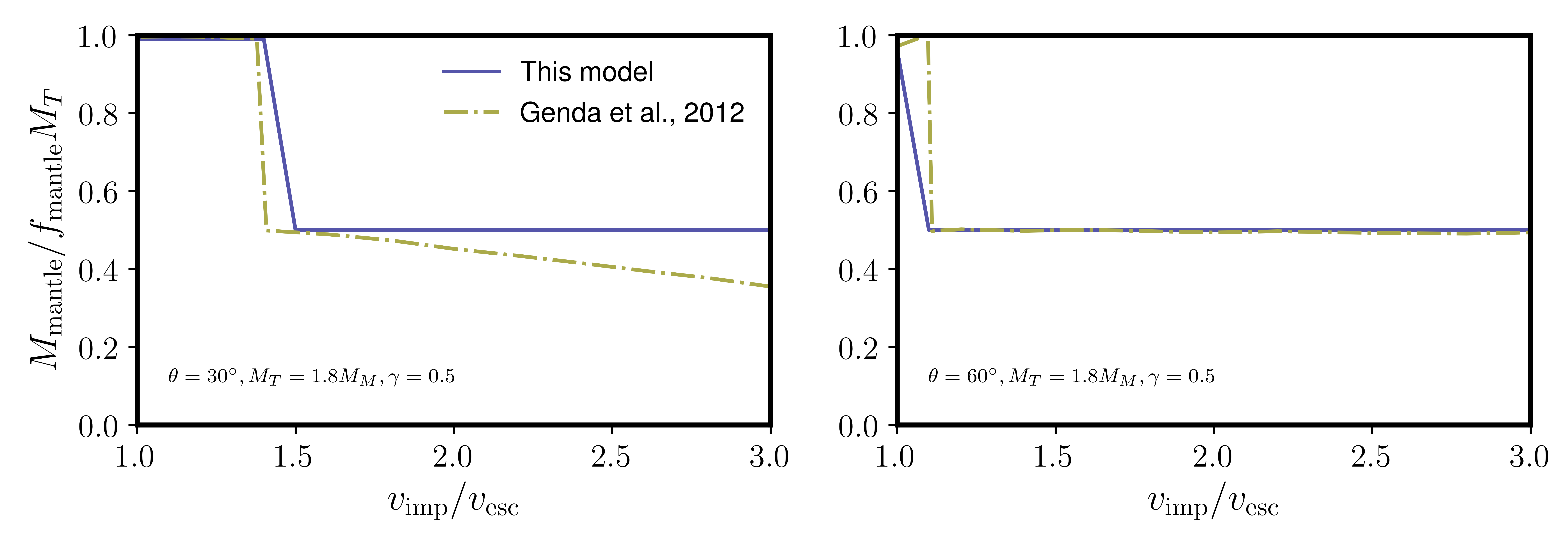}
  \end{center}
  \caption{Comparison with previous studies \citep{Gendaetal2012}.}
  \label{fig:abramov_vesc}
\end{figure*}

\begin{figure*}
  \begin{center}
    \includegraphics[scale=0.6]{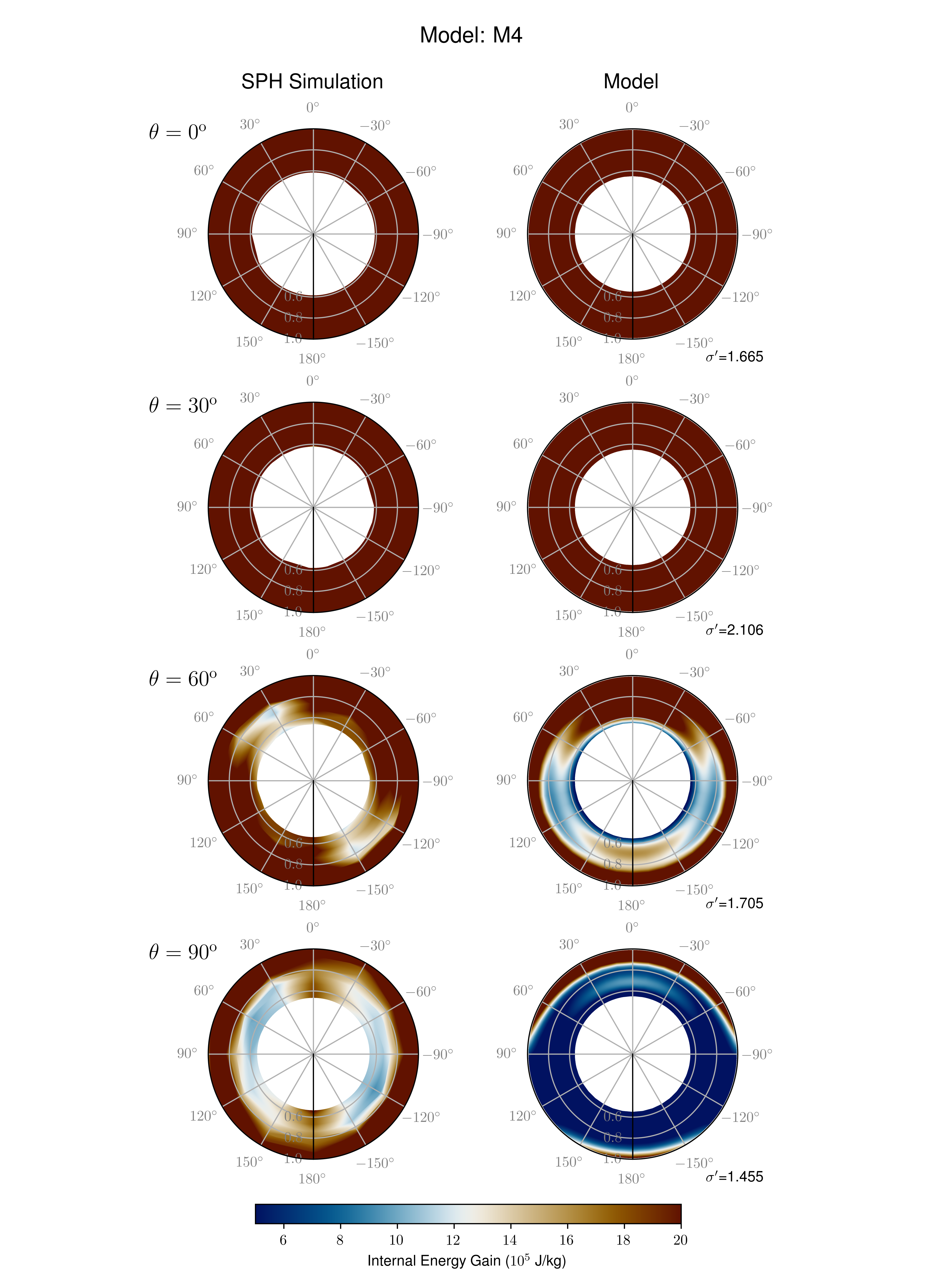}
  \end{center}
  \caption{The heat distribution of M4. The left panels correspond to the SPH simulation and the right panels correspond to our model. In this model $\gamma=0.5$ and heat distribution is not well captured at $\theta=60^\circ$ and $\theta=90^\circ$ compared to smaller $\gamma$ cases (e.g., Figures \ref{fig:M0d} and \ref{fig:M18d}).}
  \label{fig:M4d}
\end{figure*}

Figure \ref{fig:abramov_vlarge} shows pressures at the base of three melt models for the $v_{\rm imp}\geq 1.1 v_{\rm esc}$ cases (see Figure \ref{fig:rp_vesc} for the the $v_{\rm imp} = v_{\rm esc}$ cases). Figure \ref{fig:abramov_vesc} shows the comparison between our and the model by \cite{Gendaetal2012} regarding the merger condition (see Equation \ref{eq:genda}). Figure \ref{fig:M4d} shows the internal energy gain distribution for Run M4.

Some of the previous work \citep[e.g.,][]{LeinhardtStewart2012, LockStewart2017} develop and use a specific impact energy (this is  referred as $Q_R$ in the studies mentioned above) to describe the total kinetic energy involved in an impact and thus to take into account the influence of the impact angle on the heating.
This is a useful alternative parameter, and we also attempted to include this, but it did not work well for the $v_{\rm imp}=v_{\rm esc}$ cases because a significant portion of the heat is coming from energy due to merger, which is not captured in $Q_R$. The $Q_R$ model is expected to work better at $v_{\rm imp}>v_{\rm cr}$, which are more likely to end up with hit-and-run events (Equation \ref{eq:genda}).

\section{Additional output parameters}
\label{sec:additional_output_parameters}
Additional parameters are listed in Tables \ref{tb:list1_s}-\ref{tb:list3_s}. VMF is the vapor mass fraction, $T_{\rm spin}$ is the spin orbital time in hours, $I_x$ and $I_z$ are the moments of inertia along $x$ axis and $z$ axis, respectively, where the $z$ axis is perpendicular to the impact plane (the impact occurs in the $x-y$ plane, which is the same as the $\phi=0^\circ$ plane). There is practically no difference between the $x$ and $y$ axes if a post-impact body rotates. At $\theta=0^\circ$, the impact point is at $x=0$ and the post-impact body is practically not rotating. 

VMF is calculated as 
\begin{equation}
 VMF=\frac{1}{N^\prime}\sum_i^{N^\prime} 
    \begin{cases}
      0, & \text{at}\  S_i < S_{\rm liquid},    \\
  \frac{S_i-S_{\rm liquid}}{S_{\rm vapor}-S_{\rm liquid}}, & \text{at}\  S_{\rm liquid} \leq  S_{i} \leq    S_{\rm vapor},   \\
            1, & \text{at}\    S_{\rm vapor} < S_i\ {\rm and}\ P_i<P_{\rm crit}. \\
    \end{cases}
    \label{eq:mass_vimp}
\end{equation}
where $i$ represents an SPH mantle particle, $S_i$ is the entropy of the SPH particle $i$, $N^\prime$ is the total number of SPH mantle particles. $S_{\rm vapor}$ and $S_{\rm liquid}$ are the entropies of the vapor and liquid at the phase boundary (these values depend on temperature). $P$ is the pressure and $P_{\rm crit}$ is the critical pressure (2.547 GPa). The vapor mass fraction of a post-impact body is generally small for most of the simulations, but is large when the total mass is large ($M_T=26.84M_{\rm Mars}$ and $54 M_{\rm Mars}$ in M14 and M15) or the impact velocity is large ($M_T=4.23 M_{\rm Mars}, v_{\rm imp}=1.6 v_{\rm esc}$ at $\theta=0^\circ$ in M26). In these scenarios, the mantles experience almost complete melting. Thus, our assumption that vaporization does not affect the estimation of the mass of impact-induced melt seems acceptable. It should be noted, however, that VMF also depends on the choice of melt criterion (see Section \ref{sec:initT}). 

$T_{\rm spin}$ is calculated based on the angular velocity, which is estimated by dividing the angular momentum along the $z$ axis by $I_z$. At $\theta=0^\circ$, a post-impact body is not rotating, which makes $T_{\rm spin}$ large, but not infinity. This is because the boundary between a post-impact body and ejecta is not clearly defined and calculating the exact moment of inertia or $L_z$ is challenging. Nevertheless, $T_{\rm spin}$ is generally much larger at $\theta=0^\circ$ compared to the other cases.

The parameter $I_x/I_z$ is related to the oblateness of a post-impact body. When this value is close to 1, a post-impact body is close to a sphere, whereas a large deviation from 1 means that a body is more oblate. Most of the bodies have values in the range of $0.8-1$, but there are a few exceptions. For example, model M9 at $\theta=90^\circ$ (ID 40) shows $I_x/I_z=0.452$. These oblate bodies have lower pressures than calculated pressures assuming the bodies are hydrostatic, which can be seen as differences between $P_{\rm CMB, SPH}$ (46.105 GPa) and $P_{\rm CMB, Model}$ (67.127 GPa).
$P_{\rm CMB, SPH}$ is the pressure at the core-mantle boundary from an SPH simulation and $P_{\rm CMB, Model}$ is the model estimate, where no spin is considered.  In general, at small $\gamma$  ($\leq 0.1$), the difference between $P_{\rm CMB, SPH}$ and $P_{\rm CMB, Model}$ is small, but at larger $\gamma$ ($\geq 0.3$), the difference can be large due to high spin of the post-impact body.






\begin{center}
\begin{table*}[ht]
\tiny
\scalebox{1.0}{
\hfill{}
\begin{tabular}{ c c c c c c c }
\hline
 $e_0$ & $e_1$ &  $e_2$ & $e_3$ & $e_4$ & $e_5$ & $e_6$   \\
\hline
 0.18432 & 0.06338 & 0.00353 & 0.06389 & 0.10604 & -0.18243  & 0.0279  \\
\\
\hline
 $g_0$ & $g_1$ &  $g_2$ &  $k_0$  &  $k_1$    \\
\hline
 0.81590257 & 0.04083351 & -0.09894310 &  0.92329251 & 0.07644334 \\
\\
\end{tabular}}
\hfill{}
\caption{List of parameters for the internal energy gain ($e_l$, Equation \ref{eq:IE}), for the heat partitioning into the mantle ($g_l$, Equation \ref{eq:h}), and for the mantle mass for a post-impact body ($k_l$, Equation \ref{eq:mass_vesc}) at $v_{\rm imp}=v_{\rm esc}$.}
\label{tb:e}
\end{table*}
\end{center}
\begin{center}
\begin{table*}[ht]
\tiny
\scalebox{1.0}{
\hfill{}
\begin{tabular}{ c c c c c c c }
\hline
 $e_0$ & $e_1$ &  $e_2$ & $e_3$ & $e_4$ & $e_5$ & $e_6$   \\
\hline
0.01934962 & 0.045056792 & 0.11079199 & 0.17159203 & 0.14955157 &  -0.11510527 & -0.015958111  \\
\\
\hline
 $g_0$ & $g_1$ &  $g_2$     \\
\hline
0.6712941 & 0.3572683 & -0.2455803 \\
\\
\end{tabular}}
\hfill{}
\caption{Parameters are the same as Table \ref{tb:e}, but at $v_{\rm imp}\geq 1.1 v_{\rm esc}$ except that $k_0$ and $k_1$ are not used (see Equation \ref{eq:mass_vlarge}).}
\label{tb:e_largev}
\end{table*}
\end{center}

\begin{center}
\begin{table*}[ht]
\tiny
\scalebox{1.0}{
\hfill{}
\begin{tabular}{c c c c c c c c }
$\theta=0^\circ$\\
\hline
$c_0$ & $c_1$ &  $c_2$ & $c_3$ & $c_4$ & $c_5$ & $c_6$ &   $c_7$ \\
\hline
 24.353 & 4.649 & -2.640 & -154.766 & -32.733 & 11.115 & 364.404 & 82.057 \\ 
\\
\hline
 $c_8$ & $c_9$ & $c_{10}$  & $c_{11}$ & $c_{12}$ &  $c_{13}$ & $c_{14}$  \\
\hline
-15.839 & -374.516 & -88.026 & 8.886 & 142.789 & 34.572 & -1.090\\
 \hline
 \\
 \\
 $\theta=30^\circ$\\
  \hline
 $c_0$ & $c_1$ &  $c_2$ & $c_3$ & $c_4$ & $c_5$ & $c_6$ &   $c_7$ \\
\hline
 39.391 & -14.014 & -24.081 & -244.080 & 72.397 & 131.101 & 560.089 & -137.102 \\
  \\
\hline
 $c_8$ & $c_9$ & $c_{10}$  & $c_{11}$ & $c_{12}$ &  $c_{13}$ & $c_{14}$  \\
\hline
-265.118 & -562.469 & 113.394 & 237.499 & 209.616 & -34.056 & -79.446\\
 \\
 $\theta=60^\circ$\\
 \hline
$c_0$ & $c_1$ &  $c_2$ & $c_3$ & $c_4$ & $c_5$ & $c_6$ &   $c_7$ \\
\hline
  51.282 & 14.563 & 25.789 & -319.053 & -95.684 & -159.415 & 733.335 & 230.884 \\
\\
\hline
 $c_8$ & $c_9$ & $c_{10}$  & $c_{11}$ & $c_{12}$ &  $c_{13}$ & $c_{14}$  \\
\hline
  363.663 & -737.805 & -242.888 & -362.911 & 275.155 & 94.525 & 134.218\\
 \\
 $\theta=90^\circ$\\
 \hline
$c_0$ & $c_1$ &  $c_2$ & $c_3$ & $c_4$ & $c_5$ & $c_6$ &   $c_7$ \\
\hline
 64.110 & 129.937 & 125.322 & -394.726 & -783.800 & -756.950 & 897.274 & 1753.903\\
\\
\hline
 $c_8$ & $c_9$ & $c_{10}$  & $c_{11}$ & $c_{12}$ &  $c_{13}$ & $c_{14}$  \\
\hline
 1695.495 & -892.887 & -1726.224 & -1669.516 & 329.292 & 631.170 & 610.445\\

\end{tabular}}
\hfill{}
\caption{Model coefficients for $\theta=0$, 30, 60, and $90^{\rm o}$ (Equation \ref{eq:F}).}
\label{tb:coefficients0}
\end{table*}
\end{center}
\begin{center}
\begin{table*}[ht]
\tiny
\scalebox{1.0}{
\hfill{}
\begin{tabular}{c c c c c c c c c c c c}
\hline
Model & ID &  $M_T$ & $\gamma$ & $\theta(^\circ)$ & $\frac{v_{\rm imp}}{v_{\rm esc}}$  &  VMF & $T_{\rm spin}$ (hrs)  &  $I_x/I_z$  & $I_z\times 10^{-37}$  &  $P_{\rm CMB, SPH}$ (GPa) &  $P_{\rm CMB, Model}$ (GPa)\\
\hline
M0 & 1 & 1.0 & 0.1 & 0 & 1.0 & 0.003 & 1074.830 & 1.074 & 0.240 & 21.245 & 20.778\\
M0 & 2 & 1.0 & 0.1 & 30 & 1.0 & 0.002 & 8.213 & 1.006 & 0.248 & 20.186 & 20.377\\
M0 & 3 & 1.0 & 0.1 & 60 & 1.0 & 0.005 & 11.487 & 0.581 & 0.396 & 19.843 & 19.883\\
M0 & 4 & 1.0 & 0.1 & 90 & 1.0 & 0.000 & 37.781 & 0.992 & 0.210 & 18.807 & 19.225\\
M1 & 5 & 1.0 & 0.1 & 0 & 1.0 & 0.000 & 316.479 & 0.997 & 0.250 & 18.626 & 20.778\\
M1 & 6 & 1.0 & 0.1 & 30 & 1.0 & 0.001 & 7.731 & 0.957 & 0.255 & 19.404 & 20.377\\
M1 & 7 & 1.0 & 0.1 & 60 & 1.0 & 0.001 & 7.007 & 0.958 & 0.259 & 22.585 & 19.883\\
M1 & 8 & 1.0 & 0.1 & 90 & 1.0 & 0.001 & 11.082 & 0.980 & 0.229 & 20.938 & 19.225\\
M2 & 9 & 1.0 & 0.1 & 0 & 1.0 & 0.002 & 1717.294 & 1.034 & 0.245 & 18.039 & 20.539\\
M2 & 10 & 1.0 & 0.1 & 30 & 1.0 & 0.002 & 8.130 & 0.957 & 0.252 & 19.858 & 20.377\\
M2 & 11 & 1.0 & 0.1 & 60 & 1.0 & 0.003 & 17.015 & 0.958 & 0.737 & 20.732 & 20.114\\
M2 & 12 & 1.0 & 0.1 & 90 & 1.0 & 0.001 & 34.145 & 0.996 & 0.210 & 18.924 & 19.226\\
M3 & 13 & 1.03 & 0.091 & 0 & 1.0 & 0.000 & 34470.193 & 0.968 & 0.265 & 20.267 & 21.360\\
M3 & 14 & 1.03 & 0.091 & 30 & 1.0 & 0.001 & 8.834 & 0.965 & 0.265 & 20.323 & 20.946\\
M3 & 15 & 1.03 & 0.091 & 60 & 1.0 & 0.002 & 7.635 & 0.935 & 0.259 & 23.532 & 20.437\\
M3 & 16 & 1.03 & 0.091 & 90 & 1.0 & 0.001 & 41.022 & 0.792 & 0.283 & 20.595 & 19.989\\
M4 & 17 & 1.88 & 0.5 & 0 & 1.0 & 0.046 & 7530.717 & 0.784 & 0.808 & 29.276 & 35.071\\
M4 & 18 & 1.88 & 0.5 & 30 & 1.0 & 0.013 & 3.889 & 0.791 & 0.919 & 24.334 & 34.781\\
M4 & 19 & 1.88 & 0.5 & 60 & 1.0 & 0.006 & 3.641 & 0.524 & 1.153 & 20.856 & 34.320\\
M4 & 20 & 1.88 & 0.5 & 90 & 1.0 & 0.002 & 6.878 & 0.827 & 2.253 & 19.649 & 33.216\\
M5 & 21 & 3.06 & 0.032 & 0 & 1.0 & 0.006 & 7991.603 & 0.938 & 1.497 & 55.365 & 53.568\\
M5 & 22 & 3.06 & 0.032 & 30 & 1.0 & 0.005 & 23.626 & 0.984 & 1.502 & 53.332 & 53.109\\
M5 & 23 & 3.06 & 0.032 & 60 & 1.0 & 0.003 & 27.831 & 0.994 & 1.413 & 51.277 & 51.717\\
M5 & 24 & 3.06 & 0.032 & 90 & 1.0 & 0.001 & 149.991 & 0.998 & 1.389 & 50.399 & 50.653\\
M6 & 25 & 3.25 & 0.091 & 0 & 1.0 & 0.006 & 3396.710 & 1.003 & 1.650 & 56.505 & 57.345\\
M6 & 26 & 3.25 & 0.091 & 30 & 1.0 & 0.022 & 8.479 & 0.969 & 1.682 & 53.074 & 55.962\\
M6 & 27 & 3.25 & 0.091 & 60 & 1.0 & 0.025 & 7.700 & 0.950 & 1.625 & 57.411 & 54.633\\
M6 & 28 & 3.25 & 0.091 & 90 & 1.0 & 0.005 & 42.678 & 1.006 & 1.415 & 50.246 & 53.504\\
M7 & 29 & 3.25 & 0.091 & 0 & 1.0 & 0.004 & 3616.659 & 0.992 & 1.627 & 52.904 & 57.345\\
M7 & 30 & 3.25 & 0.091 & 30 & 1.0 & 0.014 & 8.465 & 1.005 & 1.598 & 59.670 & 55.962\\
M7 & 31 & 3.25 & 0.091 & 60 & 1.0 & 0.020 & 7.206 & 0.956 & 1.574 & 57.167 & 54.633\\
M7 & 32 & 3.25 & 0.091 & 90 & 1.0 & 0.003 & 43.884 & 0.991 & 1.380 & 55.430 & 53.504\\
M8 & 33 & 3.7 & 0.2 & 0 & 1.0 & 0.060 & 2253.268 & 1.017 & 2.111 & 53.751 & 63.211\\
M8 & 34 & 3.7 & 0.2 & 30 & 1.0 & 0.116 & 4.867 & 0.867 & 2.410 & 47.725 & 63.500\\
M8 & 35 & 3.7 & 0.2 & 60 & 1.0 & 0.044 & 3.583 & 0.794 & 2.435 & 40.232 & 61.809\\
M8 & 36 & 3.7 & 0.2 & 90 & 1.0 & 0.031 & 6.958 & 0.668 & 2.573 & 55.936 & 59.726\\
M9 & 37 & 4.23 & 0.301 & 0 & 1.0 & 0.223 & 3693.041 & 0.941 & 2.816 & 52.438 & 71.859\\
M9 & 38 & 4.23 & 0.301 & 30 & 1.0 & 0.157 & 3.996 & 0.826 & 3.040 & 47.533 & 71.226\\
M9 & 39 & 4.23 & 0.301 & 60 & 1.0 & 0.041 & 3.755 & 0.636 & 3.785 & 40.229 & 68.569\\
M9 & 40 & 4.23 & 0.301 & 90 & 1.0 & 0.033 & 6.002 & 0.452 & 5.012 & 46.105 & 67.127\\
\hline
\end{tabular}}
\hfill{}
\caption{Additional list of parameters for models M0-M9 (see Table \ref{tb:list1}).  VMF is the vapor mass fraction, $T_{\rm spin}$ is the spin period in hours, $I_z$ and $I_x$ are the moments of inertia along $z$ and $x$, respectively. Here, the impact plane is the $x-y$ plane and $x$ is perpendicular to the plane. The last column, $I_z$, is normalized by $10^{37}$ kg m$^2$. $P_{\rm CMB, SPH}$ is the pressure in GPa at the core-mantle boundary from an SPH simulation and $P_{\rm CMB, Model}$ is the model estimate, where no spin is considered.}
\label{tb:list1_s}
\end{table*}
\end{center}
\begin{center}
\begin{table*}[ht]
\tiny
\scalebox{1.0}{
\hfill{}
\begin{tabular}{c c c c c c c c c c  c c}
\hline
Model & ID &  $M_T$ & $\gamma$ & $\theta(^\circ)$ & $\frac{v_{\rm imp}}{v_{\rm esc}}$  &  VMF & $T_{\rm spin}$ (hrs)  &  $I_x/I_z$  & $I_z\times 10^{-37}$  &  $P_{\rm CMB, SPH}$ & $P_{\rm CMB, Model}$  \\
\hline
M10 & 41 & 5.34 & 0.091 & 0 & 1.0 & 0.021 & 2349.659 & 1.200 & 3.494 & 73.680 & 86.737\\
M10 & 42 & 5.34 & 0.091 & 30 & 1.0 & 0.081 & 8.559 & 0.955 & 3.759 & 84.112 & 87.148\\
M10 & 43 & 5.34 & 0.091 & 60 & 1.0 & 0.009 & 20.835 & 0.707 & 4.379 & 78.690 & 84.984\\
M10 & 44 & 5.34 & 0.091 & 90 & 1.0 & 0.010 & 41.717 & 0.989 & 3.075 & 78.307 & 80.913\\
M11 & 45 & 6.54 & 0.091 & 0 & 1.0 & 0.040 & 2340.101 & 0.851 & 5.231 & 105.315 & 102.928\\
M11 & 46 & 6.54 & 0.091 & 30 & 1.0 & 0.124 & 8.498 & 0.968 & 5.219 & 100.562 & 102.283\\
M11 & 47 & 6.54 & 0.091 & 60 & 1.0 & 0.013 & 20.168 & 0.780 & 5.490 & 95.361 & 101.116\\
M11 & 48 & 6.54 & 0.091 & 90 & 1.0 & 0.013 & 38.829 & 0.995 & 4.222 & 92.995 & 96.235\\
M12 & 49 & 8.94 & 0.091 & 0 & 1.0 & 0.187 & 394.894 & 1.251 & 8.647 & 99.340 & 134.408\\
M12 & 50 & 8.94 & 0.091 & 30 & 1.0 & 0.268 & 5.980 & 1.140 & 9.114 & 124.761 & 135.009\\
M12 & 51 & 8.94 & 0.091 & 60 & 1.0 & 0.036 & 14.045 & 1.016 & 7.045 & 126.205 & 131.694\\
M12 & 52 & 8.94 & 0.091 & 90 & 1.0 & 0.032 & 42.282 & 0.996 & 6.885 & 126.110 & 125.447\\
M13 & 53 & 9.43 & 0.104 & 0 & 1.0 & 0.247 & 1080.059 & 0.997 & 9.087 & 128.410 & 140.735\\
M13 & 54 & 9.43 & 0.104 & 30 & 1.0 & 0.279 & 6.502 & 0.947 & 9.311 & 130.358 & 139.485\\
M13 & 55 & 9.43 & 0.104 & 60 & 1.0 & 0.173 & 6.154 & 0.945 & 8.870 & 126.430 & 137.891\\
M13 & 56 & 9.43 & 0.104 & 90 & 1.0 & 0.033 & 31.994 & 0.954 & 7.626 & 130.372 & 133.129\\
M14 & 57 & 26.84 & 0.091 & 0 & 1.0 & 0.770 & 5883.023 & 1.000 & 43.983 & 361.563 & 364.414\\
M14 & 58 & 26.84 & 0.091 & 30 & 1.0 & 0.502 & 6.912 & 0.962 & 44.757 & 366.250 & 356.113\\
M14 & 59 & 26.84 & 0.091 & 60 & 1.0 & 0.192 & 10.440 & 0.984 & 42.126 & 371.723 & 350.714\\
M14 & 60 & 26.84 & 0.091 & 90 & 1.0 & 0.064 & 31.166 & 0.999 & 37.065 & 367.305 & 333.456\\
M15 & 61 & 53.66 & 0.091 & 0 & 1.0 & 0.964 & 9109.850 & 0.995 & 127.215 & 705.281 & 718.623\\
M15 & 62 & 53.66 & 0.091 & 30 & 1.0 & 0.757 & 6.418 & 0.959 & 129.038 & 714.908 & 721.897\\
M15 & 63 & 53.66 & 0.091 & 60 & 1.0 & 0.309 & 9.732 & 0.956 & 113.856 & 709.716 & 700.116\\
M15 & 64 & 53.66 & 0.091 & 90 & 1.0 & 0.101 & 30.287 & 0.997 & 106.793 & 712.542 & 671.301\\
\hline
\end{tabular}}
\hfill{}
\caption{
Additional list of parameters for models M10-M15 (see Table \ref{tb:list2}).}
\label{tb:list2_s}
\end{table*}
\end{center}
\begin{center}
\begin{table*}[ht]
\tiny
\scalebox{1.0}{
\hfill{}
\begin{tabular}{c c c c c c c c c c c c}
\hline
Model & ID &  $M_T$ & $\gamma$ & $\theta(^\circ)$ & $\frac{v_{\rm imp}}{v_{\rm esc}}$  &  VMF & $T_{\rm spin}$ (hrs)  &  $I_x/I_z$  & $I_z\times 10^{-37}$ & $P_{\rm CMB, SPH}$ & $P_{\rm CMB, Model}$ \\
\hline
M16 & 65 & 1.88 & 0.5 & 0 & 1.3 & 0.180 & 4627.804 & 1.090 & 0.659 & 27.245 & 35.071\\
M16 & 66 & 1.88 & 0.5 & 30 & 1.3 & 0.060 & 4.045 & 0.637 & 1.378 & 12.350 & 34.781\\
M16 & 67 & 1.88 & 0.5 & 60 & 1.3 & 0.002 & 8.878 & 0.981 & 0.222 & 20.273 & 19.752\\
M16 & 68 & 1.88 & 0.5 & 90 & 1.3 & 0.000 & 13.381 & 1.007 & 0.221 & 20.177 & 19.752\\
M17 & 69 & 3.25 & 0.091 & 0 & 1.1 & 0.015 & 4302.774 & 0.944 & 1.711 & 49.682 & 57.345\\
M17 & 70 & 3.25 & 0.091 & 30 & 1.1 & 0.029 & 8.170 & 0.964 & 1.683 & 52.251 & 55.962\\
M17 & 71 & 3.25 & 0.091 & 60 & 1.1 & 0.005 & 20.271 & 1.003 & 1.427 & 49.860 & 52.677\\
M17 & 72 & 3.25 & 0.091 & 90 & 1.1 & 0.001 & 77.743 & 0.997 & 1.387 & 49.546 & 52.677\\
M18 & 73 & 3.25 & 0.091 & 0 & 1.2 & 0.021 & 7074.261 & 1.060 & 1.640 & 49.978 & 57.345\\
M18 & 74 & 3.25 & 0.091 & 30 & 1.2 & 0.034 & 7.969 & 0.972 & 1.657 & 52.111 & 55.962\\
M18 & 75 & 3.25 & 0.091 & 60 & 1.2 & 0.005 & 21.785 & 0.993 & 1.419 & 50.367 & 52.677\\
M18 & 76 & 3.25 & 0.091 & 90 & 1.2 & 0.001 & 76.341 & 0.998 & 1.383 & 50.896 & 52.677\\
M19 & 77 & 3.25 & 0.091 & 0 & 1.3 & 0.034 & 8587.565 & 1.119 & 1.643 & 45.351 & 57.345\\
M19 & 78 & 3.25 & 0.091 & 30 & 1.3 & 0.041 & 7.952 & 0.955 & 1.657 & 52.953 & 55.962\\
M19 & 79 & 3.25 & 0.091 & 60 & 1.3 & 0.007 & 26.404 & 0.998 & 1.394 & 55.345 & 52.677\\
M19 & 80 & 3.25 & 0.091 & 90 & 1.3 & 0.001 & 201.755 & 0.998 & 1.373 & 51.628 & 52.677\\
M20 & 81 & 3.25 & 0.091 & 0 & 1.4 & 0.044 & 2848.585 & 1.128 & 1.671 & 42.670 & 57.345\\
M20 & 82 & 3.25 & 0.091 & 30 & 1.4 & 0.052 & 8.227 & 1.138 & 1.655 & 52.643 & 55.962\\
M20 & 83 & 3.25 & 0.091 & 60 & 1.4 & 0.006 & 25.944 & 0.997 & 1.398 & 51.700 & 52.677\\
M20 & 84 & 3.25 & 0.091 & 90 & 1.4 & 0.001 & 284.157 & 1.000 & 1.376 & 49.485 & 52.677\\
M21 & 85 & 3.25 & 0.091 & 0 & 1.5 & 0.056 & 6608.636 & 1.057 & 1.665 & 47.618 & 57.345\\
M21 & 86 & 3.25 & 0.091 & 30 & 1.5 & 0.062 & 7.714 & 0.994 & 1.639 & 51.533 & 55.962\\
M21 & 87 & 3.25 & 0.091 & 60 & 1.5 & 0.008 & 32.555 & 0.997 & 1.396 & 50.688 & 52.677\\
M21 & 88 & 3.25 & 0.091 & 90 & 1.5 & 0.001 & 400.109 & 0.997 & 1.376 & 49.665 & 52.677\\
M22 & 89 & 3.25 & 0.091 & 0 & 2.0 & 0.340 & 3269.097 & 1.081 & 1.719 & 42.686 & 57.345\\
M22 & 90 & 3.25 & 0.091 & 30 & 2.0 & 0.106 & 10.451 & 0.965 & 1.524 & 48.799 & 52.677\\
M22 & 91 & 3.25 & 0.091 & 60 & 2.0 & 0.008 & 64.683 & 1.005 & 1.395 & 47.175 & 52.677\\
M22 & 92 & 3.25 & 0.091 & 90 & 2.0 & 0.000 & 44.092 & 1.000 & 1.383 & 49.390 & 52.677\\
M23 & 93 & 4.23 & 0.301 & 0 & 1.1 & 0.214 & 1939.650 & 0.715 & 3.083 & 48.094 & 71.859\\
M23 & 94 & 4.23 & 0.301 & 30 & 1.1 & 0.203 & 4.142 & 0.811 & 3.134 & 46.267 & 71.226\\
M23 & 95 & 4.23 & 0.301 & 60 & 1.1 & 0.010 & 11.238 & 0.987 & 1.454 & 49.669 & 52.677\\
M23 & 96 & 4.23 & 0.301 & 90 & 1.1 & 0.003 & 18.179 & 0.979 & 1.401 & 50.451 & 52.677\\
M24 & 97 & 4.23 & 0.301 & 0 & 1.2 & 0.290 & 2915.583 & 1.374 & 2.471 & 41.815 & 71.859\\
M24 & 98 & 4.23 & 0.301 & 30 & 1.2 & 0.231 & 4.195 & 0.859 & 2.989 & 42.585 & 71.226\\
M24 & 99 & 4.23 & 0.301 & 60 & 1.2 & 0.008 & 19.433 & 1.073 & 1.406 & 48.171 & 52.677\\
M24 & 100 & 4.23 & 0.301 & 90 & 1.2 & 0.002 & 136.449 & 1.007 & 1.383 & 53.551 & 52.677\\
M25 & 101 & 4.23 & 0.301 & 0 & 1.3 & 0.436 & 1589.596 & 0.870 & 2.845 & 48.258 & 71.859\\
M25 & 102 & 4.23 & 0.301 & 30 & 1.3 & 0.313 & 4.875 & 0.753 & 3.678 & 37.979 & 71.226\\
M25 & 103 & 4.23 & 0.301 & 60 & 1.3 & 0.010 & 19.620 & 1.103 & 1.407 & 49.426 & 52.677\\
M25 & 104 & 4.23 & 0.301 & 90 & 1.3 & 0.002 & 879.278 & 1.012 & 1.392 & 51.957 & 52.677\\
M26 & 105 & 4.23 & 0.301 & 0 & 1.6 & 0.571 & 1295.750 & 0.720 & 2.661 & 31.144 & 71.859\\
M26 & 106 & 4.23 & 0.301 & 30 & 1.6 & 0.179 & 20.967 & 0.985 & 2.230 & 43.662 & 52.677\\
M26 & 107 & 4.23 & 0.301 & 60 & 1.6 & 0.012 & 15.578 & 0.963 & 1.421 & 52.737 & 52.677\\
M26 & 108 & 4.23 & 0.301 & 90 & 1.6 & 0.002 & 419.454 & 1.008 & 1.389 & 49.664 & 52.677\\
\hline
\end{tabular}}
\hfill{}
\caption{ Additional list of parameters for models M16-M26 (see Table \ref{tb:list3}).}
\label{tb:list3_s}
\end{table*}
\end{center}

\end{document}